\documentstyle[prl,twocolumn,aps,epsf]{revtex}

\begin{document}
 \tolerance 50000

\draft

\twocolumn[\hsize\textwidth\columnwidth\hsize\csname@twocolumnfalse\endcsname

\title{Diagonal Ladders: A New Class of Models for
Strongly Coupled Electron Systems} 

\author{G. Sierra$^{1}$,  
M.A. Mart{\'\i}n-Delgado$^{2}$,
S.R. White$^{3}$, D.J. Scalapino$^{4}$, J. Dukelsky$^{5}$
 } 
\address{ 
$^{1}$Instituto de Matem{\'a}ticas y F{\'\i}sica Fundamental, C.S.I.C.,
Madrid, Spain. 
\\ 
$^{2}$Departamento de
F{\'\i}sica Te{\'o}rica I, Universidad Complutense. Madrid, Spain.
 \\
$^{3}$Department of Physics and 
Astronomy, University of California, Irvine,CA 92697 \\
$^{4}$Department of Physics,
University of California,
Santa Barbara, CA 93106 \\
$^{5}$Instituto de Estructura de la Materia, C.S.I.C.,Madrid, Spain.  
}

\maketitle 

\begin{abstract} 
\begin{center}
\parbox{14cm}{ We introduce a class of models defined on ladders
with a diagonal structure generated by $n_p$ plaquettes.  The case
$n_p=1$ corresponds to the necklace ladder and has remarkable
properties which are studied using DMRG and recurrent variational
ansatzes.  The AF Heisenberg model on this ladder is equivalent to the
alternating spin-1/spin-$\frac{1}{2}$ AFH chain which is known to have
a ferrimagnetic ground state (GS). For doping 1/3 the GS is a fully
doped (1,1) stripe with the holes located mostly along the principal
diagonal while the minor diagonals are occupied by spin singlets.
This state can be seen as a Mott insulator of localized Cooper pairs
on the plaquettes.  A physical picture of our results is provided by a
$t_p-J_p-t_d$ model of plaquettes coupled diagonally with a hopping
parameter $t_d$.  In the limit $t_d \rightarrow \infty$ we recover the
original $t-J$ model on the necklace ladder while for weak hopping
parameter the model is easily solvable. The GS in the strong hopping
regime is essentially an ``on link" Gutzwiller projection of the weak
hopping GS.  We generalize the $t_p-J_p-t_d$ model to diagonal ladders
with $n_p >1$ and the 2D square lattice. We use in our construction
concepts familiar in Statistical Mechanics such as medial graphs and
Bratelli diagrams.}

\end{center}
\end{abstract}

\pacs{
\hspace{2.5cm} 
}

\vskip2pc]
\narrowtext

\section{Introduction}

Ladders provide a class of interesting
theoretical models for studying 
the behavior of strongly correlated electron
systems. Besides representing simplified models
for actual materials, ladders offer a possible way
of interpolating between 1 and 2 
spatial dimensions  with the hope that
they will yield  insights into
the physics of 2D systems, such as 
the $CuO_2$ planes of the cuprates (for a review see \cite{DR}).

It has been found that 
ladders exhibit quite different behavior
depending on whether the number of legs 
$n_\ell$ is even or odd. 
Antiferromagnetic spin ladders
with $n_\ell$ odd are gapless with 
spin-spin correlation functions decaying algebraically, while 
even leg ladders are gapped with a finite 
spin correlation length. 
Upon doping, these
two types of ladders also 
behave differently concerning the existence of
pairing of holes or spin-charge separation.
In the limit where the number of legs goes to infinity
the spin gap of the even spin ladders vanishes exponentially fast, 
in agreement with the gapless nature of the 2D magnons \cite{gap}.
On the other hand, the antiferromagnetic
long range order (AFLRO) characteristic 
of the 2D antiferromagnetic Heisenberg  (AFH) model can be 
more naturally attributed to  the 
quasi long range order
of the odd leg ladders. 
It thus seems that one has to combine different properties
of the even and odd, doped and undoped 
ladders in order to arrive at a consistent 
picture of the 2D cuprates.
Ladder systems are  
sufficiently
interesting on their own to deserve detailed studies,
in addition there are a variety of materials 
which contain weakly coupled arrays of ladders\cite{array}.

In this paper we study of a class of ladders
characterized by a diagonal structure which provides an alternative
to the aforementioned route to 2D. 
We shall call these
objects diagonal ladders in order 
to distinguish them  from the more familiar 
rectangular shaped ones. Diagonal ladders are  
labelled  by an integer $n_p=1,2, \dots$ which
gives the number of elementary plaquettes needed to generate
the entire structure.
The first member
of this family, i.e. $n_p=1$, is also known as the necklace ladder
and it consists of a collection of $N$ plaquettes joined along a common
diagonal. In this paper we shall focus on the necklace ladder,
although the other cases will also be briefly considered.

The original motivation of this work was to 
understand the 
fully doped stripes in the (1,1) direction
that have been observed experimentally
in materials like La$_{1-x}$Sr$_{x}$NiO$_4$\cite{Tran},
in Hartree-Fock studies of $t-J$ and Hubbard
models\cite{hfdomain},
and numerically in density matrix renormalization
group\cite{dmrg} (DMRG) studies of the $t-J$ model \cite{WS-st}. 
The simplest possible toy model
of this type of stripes is provided by a necklace ladder
with a hole doping equal to 1/3. 
As we shall see this doping plays an important role in our
work.

Lattices similar to the diagonal ladders, but with additional
one-electron hopping terms along the major and minor diagonals
of each plaquette, and with $J/t=0$, have been
solved exactly for certain fillings\cite{brandt,tasaki,giesekus}. 
Giesekus\cite{giesekus}
has shown that for the corresponding version of the necklace 
ladder, in the case when
all of the one-electron hopping terms are equal and the hole
doping is set to $x=1/3$, the model has a
short range RVB ground state, in which the static correlations
exhibit an exponential decay and the dynamic correlation
functions exhibit a gap in their spectral densities. 

Let us also note in passing that diagonal ladders have recently
appeared as constituent parts of some interesting materials like
Sr$_{0.4}$Ca$_{13.6}$Cu$_{24}$O$_{41.84}$ 
known for its superconducting properties at high pressure \cite{uehara}

The organization of the paper is as follows. In Section II
we define the diagonal ladders from a geometrical viewpoint
and compare them with the more familiar ladder structures.
In Section III we study the AF Heisenberg model of the
necklace ladder. In Section IV we study the $t-J$ model
on the necklace ladder and show the conservation of the
parity of the  plaquettes. In Section V 
we present the ground state (GS) structure  of
a necklace $t-J$ ladder with 7 plaquettes, 
obtained with the
DMRG and recurrent variational ansatz (RVA) methods. 
In Section VI we study in more
detail the structure of the GS at doping 1/3.  In Section VII
we introduce a generalized $t-J$ model on an enlarged necklace
ladder, called the $t_p-J_p-t_d$ model,
and use it to give  a physical picture of the results of Sections
V and VI. In Section VIII we define the $t_p-J_p-t_d$
model on  diagonal ladders with more than one plaquette
per unit cell and on  the 2D square lattice. In Section IX we state
our conclusions. There are three appendices which give the technical
details concerning the RVA calculations (Appendix A), the 
complete spectrum  of the $t-J$ Hamiltonian on a plaquette (Appendix B)
and a plaquette  derivation 
of the equivalence between the  spin 1 AKLT state of a
chain and the dimer-RVB state of the 2-leg AFH ladder (Appendix C). 

\begin{figure}
\hspace{-0.8cm}
\epsfxsize=12cm \epsffile{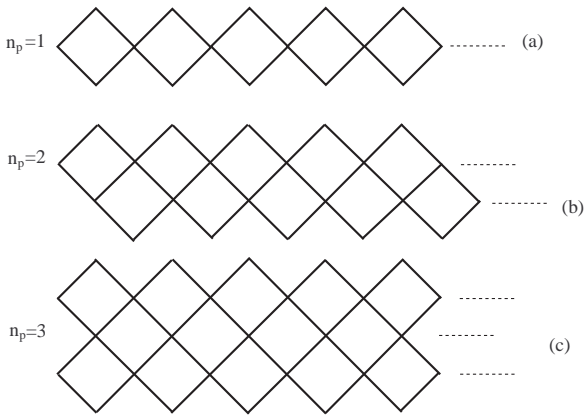}
\narrowtext
\caption[]{Examples of diagonal ladders with 
a number of plaquettes $n_p=1,2,3$ in the unit
cell.}
\label{fig1} 
\end{figure}
\noindent

\section{Geometry of Diagonal Ladders}

A diagonal ladder can be characterized by  the number of plaquettes 
$n_p$ of the unit cell and the number $N$ of  
these cells. In Fig. 1 we show 
diagonal ladders with $n_p=1,2$ and 3. There are $n_p+2$
sites per unit cell. Assuming open boundary conditions the total
number of sites is then given by $N_s = (n_p+2) N + n_p$.  

\begin{figure}
\hspace{-0.0cm}
\epsfxsize=9cm \epsffile{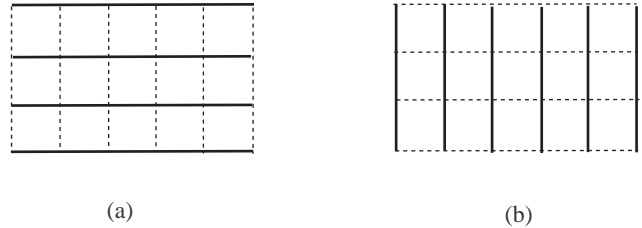}
\narrowtext
\caption[]{(a)Weak coupling (dashed lines) regime in rectangular ladders.
(b)Strong couping (dashed lines) regime in rectangular ladders.}
\label{fig2} 
\end{figure}
\noindent

\begin{figure}
\hspace{-0.0cm}
\epsfxsize=9cm \epsffile{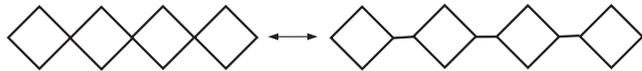}
\narrowtext
\caption[]{Plaquette construction (right) of a diagonal
ladder (left).}
\label{fig3} 
\end{figure}
\noindent

Rectangular ladders can be seen either as a collections
of legs coupled along the rungs  or as collections of rungs coupled
along the legs (see Fig. 2). This geometric feature 
is the basis of the weak coupling and strong coupling approaches
to the various physical models defined on ladders. 
Thus for example the Heisenberg model on the
$n_\ell$-leg ladder is usually  defined with an exchange
coupling constant $J_\parallel$ along the legs and an exchange
coupling constant $J_\perp$ along the rungs. 
The weak and strong coupling limits correspond to the cases
where $J_\parallel >> J_\perp$ and 
$J_\parallel << J_\perp$ respectively.

On the other hand 
diagonal ladders do not
admit such a simple construction.
The most natural interpretation is to regard them
as collections of plaquettes joined along their 
common diagonal (see Fig. 3). The trouble with this construction
is that it does not preserve the number of sites!
Indeed one has to
fuse the points on the principal diagonal of the plaquettes
before getting the actual
necklace structure. 
We shall resolve this problem in Section VII on
physical grounds.

\section{The spin necklace ladder}

Let us begin by considering  the AFH model on the
necklace ladder of Fig. 1(a). The Hamiltonian of the model is
simply,

\begin{equation}
H= J \;\sum_{< i,j >} {\bf{S}}_i \cdot {\bf{S}}_j
\label{1}
\end{equation}

\noindent
where $J$ is a positive exchange coupling constant and
the sum runs over all links $< i,j >$ of
the ladder. We shall label the sites of the $n^{\rm th}$ 
plaquette as in Fig. 4. The Hamiltonian (\ref{1}) then 
becomes

\begin{equation}
H= J \sum_{n=1}^N \;\; ( {\bf{S}}_{1,n} + {\bf{S}}_{2,n} )
\cdot (  {\bf{S}}_{3,n} +   {\bf{S}}_{3,n-1})
\label{2}
\end{equation}

\noindent where ${\bf {S}}_{a,n}$ is a spin 1/2 operator
acting at the $a=1,2,3$ position of the $n^{\rm th}$ plaquette.
Eq.(\ref{2}) implies that $H$ depends on the spins of the minor
diagonals through their sum

\begin{equation}
{\bf {S}}_{12,n} \equiv {\bf {S}}_{1,n} +  {\bf {S}}_{2,n}
\label{3}
\end{equation}

\begin{figure}
\hspace{-0.0cm}
\epsfxsize=9cm \epsffile{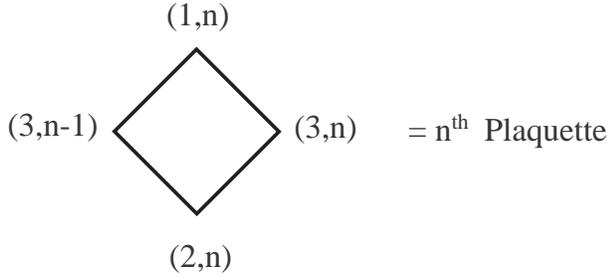}
\narrowtext
\caption[]{Diagonal coordinates for a single plaquette.}
\label{fig4} 
\end{figure}
\noindent

At this stage we are free to choose the spins of 
the $n^{\rm th}$ diagonal
in the singlet ($S_{12,n} =0$) or triplet
( $S_{12,n} =1$) representations. In the latter case the Hamiltonian
(\ref{2}) becomes entirely equivalent 
to that of an alternating spin-1/spin-$\frac{1}{2}$ chain. Choosing a 
singlet on the minor diagonal of
a given  plaquette amounts to 
adding a spin zero impurity  on the corresponding spin-1 site in the 
alternating chain, which therefore breaks into two disconnected
pieces. The net result is that the spin necklace ladder in fact
describes alternating spin-1/spin-$\frac{1}{2}$ 
chains with all possible sizes.

Fortunately, the alternating spin-1/spin-$\frac{1}{2}$ chain  
has been the
subject of several studies concerning the ground state (GS),
excitations, thermodynamic and magnetic properties \cite{Pati,Kole,alter}.
The GS turns out to be ferrimagnetic with total spin
given by $s_G= N/2$ where N is the number of unit cells 
of the chain. 
There are gapless excitations to states with spin $s_G -1$ and
gapped excitations to states with spin $s_G+1$. In spite of the
existence of gapless excitations, the chain has a finite correlation
length $\xi \sim 1 $, defined from the exponential decay of 
the spin-spin correlation function 
$< {\bf{S}}_{i} \cdot {\bf{S}}_j >$
after subtraction of the ferrimagnetic long range
contribution. These results have been obtained by a combination
of spin-wave, variational and DMRG techniques, with very satisfactory
quantitative and qualitative
agreement among them \cite{Pati,Kole}. 

\begin{figure}
\hspace{-0.0cm}
\epsfxsize=8cm \epsffile{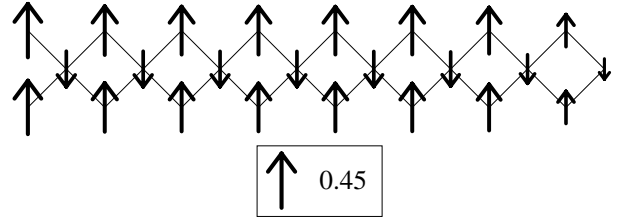}
\narrowtext
\caption[]{DMRG results for the spin configuration
for the G.S. of an undoped $8\times 3$ necklace
ladder. The state has total spin $s_G=4$.
The length of the arrow is proportional to $\langle S_z \rangle$,
according the scale in the box.}
\label{fig5} 
\end{figure}
\noindent

We have confirmed some of these properties by
applying DMRG and variational methods 
to the spin necklace ladder. In Fig. 5 we present a snapshot
of the spin configurations of the GS of an $8 \times 3$ ladder,
obtained with the DMRG, which has total spin $s_G =4$.
We find that  
the mean value of the spins near the center of the system
are given by 
$< S^z_{1,n} > = < S^z_{2,n} > =0.396$ and 
$< S^z_{3,n} > =-0.292$ 
in agreement  with the results of Ref. \cite{Pati} namely  
$< S^z_{1,n} > = 0.39624$ and  
$< S^z_{3,n} > = -0.29248$. Also using variational RVA
methods we have found $< S^z_{1,n} > = 0.4160$ and  
$< S^z_{3,n} > = -0.2893$. 

\begin{figure}
\hspace{-0.0cm}
\epsfxsize=9cm \epsffile{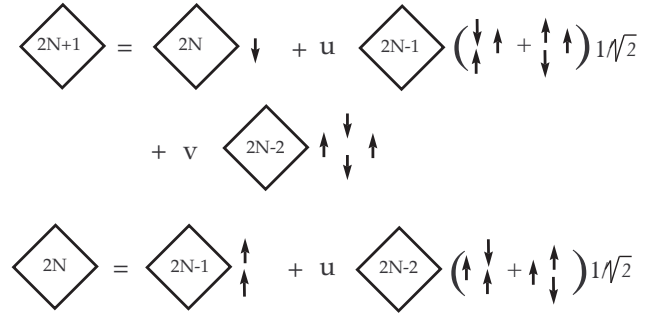}
\narrowtext
\caption[]{Diagrammatic representation of the Recurrent
Relations generating the G.S. of an undoped
necklace ladder using the variational RVA method (see Appendix A).}
\label{fig6} 
\end{figure}
\noindent

The existence of a very short
correlation length 
suggests that the ferrimagnetic GS is an adiabatic
deformation of 
the N{\'e}el state, which can be described by
a short range variational state. 
References
\cite{Pati,Kole} propose several
variational matrix product states \cite{MP}. 
It is more convenient for our purposes to use
the Recursive Variational Approach (RVA) 
of references \cite{RVA1,RVA2}, in order
to deal with doped and undoped cases on equal footing.
The GS of a ladder of length $N$ is built up
from the states 
with lengths $N-1, N-2$ and eventually $N-3$ if $N$ is odd.
The GS thus generated is a $3^{\rm rd}$ order RVA state. 
In Fig. 6 we show a diagrammatic representation of the
corresponding 
recurrence relations (we leave for Appendix A the technical
details). The GS energy per site of the associated alternating chain
that we obtain is given by $-0.7233 J$, which is to be compared with
the extrapolated DMRG results  $-0.72704 J$
or the spin-wave value $-0.718 J$ of Pati et. al. \cite{Pati}.

\section{The $\lowercase{t}-J$ model on the necklace ladder}

The Hamiltonian of the $t-J$ model is given by

\[
H_{t, J} = {\cal P}_G\left( J \; \sum_{< i,j >} \;\;
(  {{\bf S}}_i \cdot {{\bf{S}}_j} - \frac{1}{4} \;
n_i \; n_j )  \right) {\cal P}_G
\]
\begin{equation}
-{\cal P}_G\left(  t \; \sum_{ < i,j >,s}
\;( c^\dagger_{i,s} \; c_{j,s} + h. c. ) \right) {\cal P}_G
\label{4}
\end{equation}

\noindent
where the $c_{i,s}\; (\; c^\dagger_{i,s}\;) $ is the 
electron destruction (creation) operator
for site $i$ and spin $s$, $n_i$ is the occupation
number operator, 
and ${\cal P}_G$ is the Gutzwiller
projection operator which forbids doubly occupied sites.
The density-density and kinetic terms in 
(\ref{4}) can be written in a form similar
to (\ref{2}) for the exchange terms.
This suggest that there will also be a ``decoupling"
of degrees of freedom associated with the transverse diagonals.
The simplest way to see how this decoupling works
is as follows.

For the necklace $t-J$ ladder, there is a parity
plaquette conservation theorem\cite{giesekus}:
the  $t-J$ Hamiltonian on a necklace ladder 
commutes with every  graded permutation 
operator $P_n$ associated with the minor diagonal of 
the $n^{\rm th}$  plaquette. Here the permutation
operator 
$P_n (n=1, \dots, N)$ is defined by its
action on the fermionic operators, which is trivial
at all the sites except for those on the minor diagonal 
of the $n^{\rm th}$ plaquette where it acts as

\begin{eqnarray}
& P_n \; c_{(1,n),s} \; P_n^\dagger = \; c_{(2,n),s} & \nonumber \\
& P_n \; c_{(2,n),s} \; P_n^\dagger = \; c_{(1,n),s} & \label{5} 
\end{eqnarray}

\noindent
Of course the spin and the density number operators
at the sites $(1,n)$ and $(2,n)$ are also interchanged 
under the action of $P_n$. The above theorem is the
statement that $P_n$ commutes with $H_{t,J}$, Eq.(4),
for all $n$,

\begin{equation}
[H_{t,J}, \; P_n] = \; 0 \;\; {\rm for}\;\; \; n= 1, \dots, N
\label{6}
\end{equation}

\noindent
and  can be easily proved. Eq.(\ref{6}) is not special
to the $t-J$ Hamiltonian, since any other lattice
Hamiltonian having the permutation symmetry 
between the two sites on the minor diagonal of every plaquette would share
this same property.

The immediate consequence of (\ref{6}) is that we
can simultaneously diagonalize the Hamiltonian  $H_{t, J}$
and the whole collection of 
permutations operators $P_n$,  whose possible eigenvalues 
are given by 
$\epsilon_n = \pm 1$. The latter fact is a consequence of the Eq.

\begin{equation}
P_n^2 = 1 
\label{7}
\end{equation}

Letting $\epsilon_n$ denote the parity of the 
$n^{\rm th}$ plaquette,
the 9 possible states  associated 
with  the minor diagonal of a   plaquette 
can be  classified according to their 
parity, i.e. $\epsilon_n = 1$ for even parity states 
and $\epsilon_n= -1$ for odd-parity states (see Table 1).

\begin{center}
\begin{tabular}{|l|c|c|} 
\hline
 & State  & $\epsilon$ \\ 
\hline
2  holes & $|0 \rangle$ & 1  \\
bonding  & $(c^\dagger_{1,s} + 
c^\dagger_{2,s} ) |0\rangle $ & 1 \\
singlet & $\Delta^\dagger_{1,2} |0\rangle$ & 1 \\
\hline
antibonding & $(c^\dagger_{1,s} -
c^\dagger_{2,s} ) |0\rangle$ & $-1$ \\
triplet  &$( c_{1,s}^\dagger 
c_{2,s'}^\dagger + s \leftrightarrow s')
|0\rangle $ & $-1$ \\
\hline
\end{tabular}
\end{center}
\begin{center}
Table 1. Classification of the states of the minor diagonal of
a plaquette according to their parity. 
$\Delta_{1,2}^\dagger = 
\left( c_{1,\uparrow}^\dagger c_{2,\downarrow}^\dagger
+ c_{2,\uparrow}^\dagger c_{1,\downarrow}^\dagger \right)/\sqrt{2}$
is the pair field operator. 
\end{center}

The Hilbert space ${\cal H}_{\rm necklace}$ of the $t-J$ model
can be split into
a direct sum of subspaces
${\cal H}_{\bf \epsilon}$ 
classified by the parity of their plaquettes, 
$ {\bf \epsilon}=  \{ \epsilon_n \}_{n=1}^N$,
namely

\begin{equation}
{\cal H}_{\rm necklace} = \oplus_{{\bf \epsilon}} \;  
{\cal H}_{\bf \epsilon}
\label{8}
\end{equation}

\noindent
Every subspace 
${\cal H}_{\bf \epsilon}$ is left invariant under the 
action of the $t-J$ Hamiltonian (\ref{4})
which can therefore be  projected 
into an ``effective"  Hamiltonian $H_{t,J}(\epsilon)$.
In the previous section we have already seen an example
of this type of decoupling phenomena. Indeed the
alternating spin-1/spin-$\frac{1}{2}$ chain corresponds
precisely to the case where all the plaquettes are odd
and there are no holes. If holes are allowed then one has 
to consider, in addition to the triplets,
the antibonding states on the odd plaquettes. Hence there
are a total of 5 states at each site of the 
``effective" alternating chain
associated with the minor diagonal of the odd plaquettes.

On the other hand, if the parity of the plaquette is even, then
the corresponding site in the chain has 
4 possible states which can be put  into one-to-one
correspondence with those of a Hubbard  model as 
follows,

\begin{equation}
\begin{array}{rcl} {\rm Even}\;
{\rm diagonal}& \longleftrightarrow  &{\rm  Hubbard} \;{\rm site} 
\\
{\rm empty (2\; holes)} &  \longleftrightarrow & {\rm empty (1\; hole)} \\
{\rm bonding}\; {\rm states} \;(\uparrow,\downarrow) 
&  \longleftrightarrow & 
{\rm single }\; {\rm occupied} \; {\rm state} \; (\uparrow,\downarrow) \\
{\rm singlet} \; {\rm state} &  \longleftrightarrow 
& {\rm doubly}\; {\rm  occupied }\; {\rm state} \\
\end{array}
\label{9}
\end{equation}

In this case the Hamiltonian of the model is a $t-J$ Hamiltonian
with hopping parameter equal to $\sqrt{2} t$ and with the same 
exchange and density-density couplings. There is no Hubbard $U$
term.

In summary, theorem (\ref{6}) implies that
the necklace ladder is in fact equivalent to a huge collection
of alternating chains models where half of the sites
are $t-J$ like while the other half may be either spin 1
or spin-$\frac{1}{2}$ antibonding states 
for odd parity, or Hubbard like, for even parity.

It is beyond the scope of the present paper to study
such an  amazing variety of chain models disguised 
in the innocent looking  necklace ladder. 
Instead,  a more reasonable strategy is to ask 
for the values of the total spin $S$ and 
parity $\epsilon$ which give the absolute minimum of 
the GS energy, keeping  fixed the  values of 
the number of plaquettes $N$, the number of holes $h$
and the ratio $J/t$, i.e.

\begin{equation}
E_0(N,h,S_{{\rm min}},{\bf \epsilon}_{\rm min}, J/t)
\leq  E_0(N,h,S,{\bf \epsilon}, J/t), \;\;\; \forall S, {\bf \epsilon}, 
\label{10}
\end{equation}

\noindent
Even this question is not easy to answer with full generality. 
However we shall study a few cases which suggests a general
pattern for the behavior of the spin and parity as functions
of doping.

\section{DMRG and RVA results 
on the necklace ladder}

We shall concentrate on the case of
a necklace ladder with 7 plaquettes
and open BCs, which will allow us to present
in a simple manner 
the basic features of the GS for various spins ($S$)
and dopings ($h$).
The values of the coupling constants
are fixed to $t=1$ and $J=0.35$.

In Table 2 we show, for several pairs $(h,S)$ 
the parities of the plaquettes, the total GS energy
computed with the DMRG and the RVA methods. This table
also lists 
the label of the corresponding figures showing DMRG results for
the hole and spin densities of the corresponding state.

\begin{figure}
\hspace{-0.0cm}
\epsfxsize=9cm \epsffile{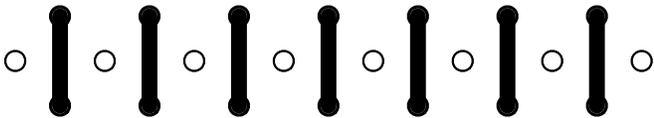}
\narrowtext
\caption[]{Pictorical representation of the most
probable state for doped $x=1/3$ necklace ladder
of dimension $8\times 8$.
Blank circles denote holes and vertical solid lines
represent valence bond states (Case (8,0)).}
\label{fig7} 
\end{figure}
\noindent

The RVA results have been derived from an inhomogeneous 
recurrence variational ansatz 
(see Appendix A). 
 As in the latter cases we start
from a state, hereafter called ``classical",  which 
is considered to be the most important configuration
present in the actual GS. 
Next we  include the local quantum 
fluctuations around the classical state.

\begin{figure}
\hspace{-1cm}
\epsfysize=9cm 
\epsfxsize=9cm \epsffile{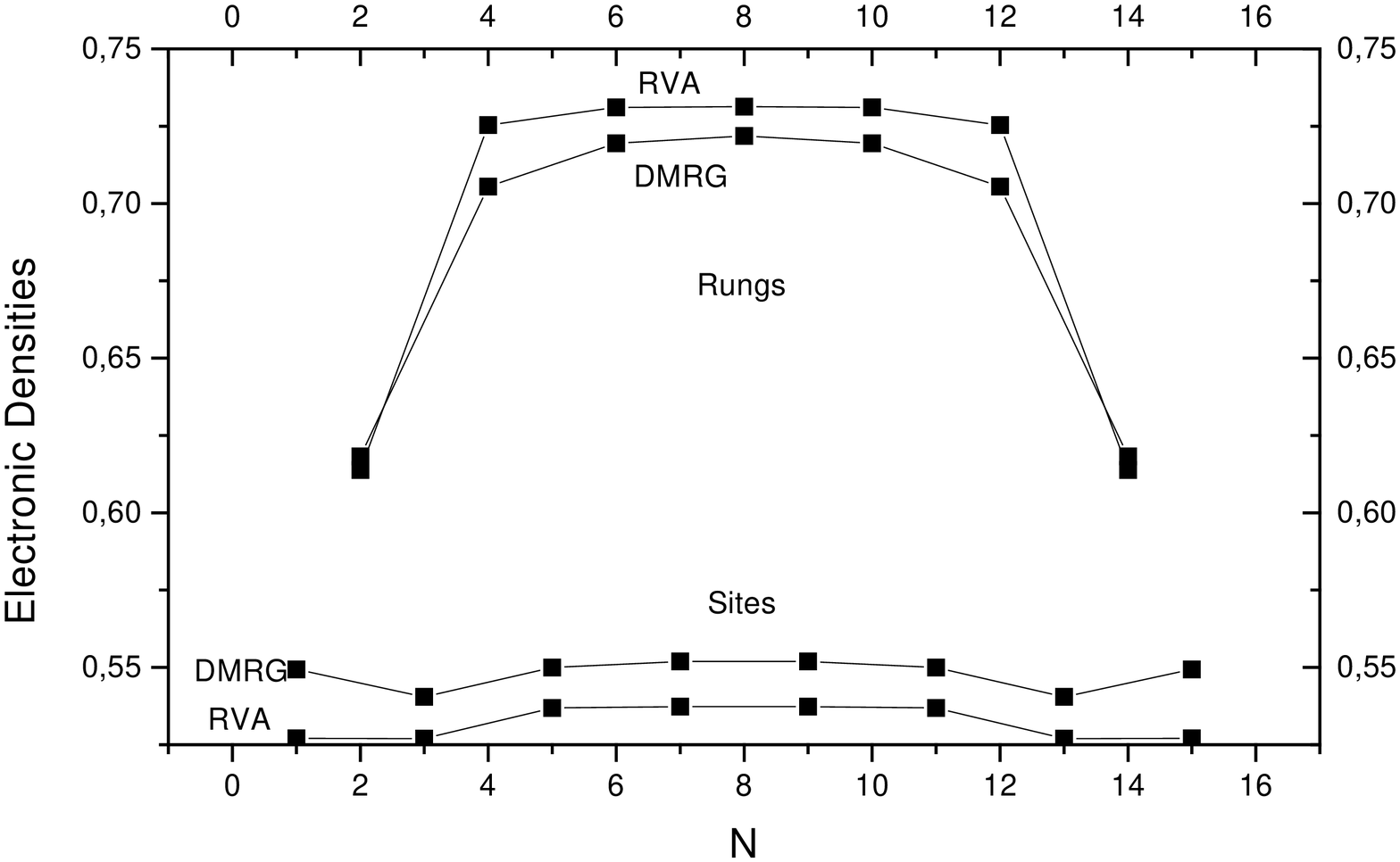}
\narrowtext
\caption[]{Electronic densities for a necklace ladder as depicted in Fig. 7.
It corresponds to 7 plaquettes and 8 holes (doping $x=1/3$). Below are shown the
DMRG and RVA values for the sites in the main diagonal, while above are shown the
values on the rungs corresponding to subdiagonals.}
\label{fig8} 
\end{figure}
\noindent

This is done for a whole set of
``classical" states  having the same number of plaquettes, 
holes and  $z$-component of  
spin. As discussed in the appendix, 
the classification of the ``classical"
states is achieved by means of paths in a Bratelli
diagram generated by folding and repeating 
the Dynkin diagram
of the  exceptional group Lie $E_6$. The six points of 
$E_6$ are in one-to-one  correspondence with 6 different states
on the necklace ladder, while the links of $E_6$ are
nearest neighbor  constraints derived  on the basis
of the DMRG results in the region $ 0 \leq x \leq 1/3$. The Bratelli 
construction gives us  a systematic
way to explore the GS manifold in the underdoped region $x \leq 1/3$.
The overdoped region has to be studied with delocalized RVA states
as discussed in references \cite{RVA1,RVA2}.

 \begin{center}
\begin{tabular}{|c|c|c|c|c|c|} 
\hline
h & S  & ($\epsilon_1, \dots, \epsilon_7)$ & Figure &
$E^{\rm DMRG}_0$ & $E^{\rm RVA}_0$  \\ 
\hline
8 & 0 &  $ (+++++++)$ & 7 and 8 & -16.554153 & -16.33996 \\
8 & 1 &  $ (+++-+++)$ & 9 &       -16.284855 & -16.00358 \\
7 & 1/2 &$ (++++++-)$ & 10 & -15.489511 & -15.18141 \\
6 & 0 &  $ (-+++++-)$ & 11 & -14.424805 & -14.02286 \\
6 & 1 &  $ (-+++++-)$ & 12 & -14.424798 & -14.02286 \\
9 & 1/2 &$ (+++++++)$ & 13 & -16.746112 & - \\
10& 0 &  $ (+++++++)$ & 14 & -16.927899 & - \\
10& 1 &  $ (+++++++)$ & 15 & -16.718476 & - \\
\hline
\end{tabular}
\end{center}
\begin{center}
Table 2. DMRG and RVA total energies for a 7-plaquette necklace ladder with $h$ holes and 
total spin $S_z$. The string of epsilons is the pattern of parities in the subdiagonals
(rungs). 
\end{center}

Let us comment on the DMRG results shown in the figures.

\begin{itemize}

\item {\bf Case (8,0):}
This is the most interesting case and 
it corresponds to one hole per site 
along  the principal diagonal. In the infinite
length limit this state has doping $x=1/3$.
For this reason we shall hereafter call this
state the $x=1/3$ state.
Figure 7 shows the 
most probable configuration which occurs when 
the holes occupy the principal diagonal of the ladder
and the spins form perfect singlets along 
the minor diagonals. The latter fact implies that
all the plaquettes are even (see Table 1). 
Figure 8 shows the electronic density along the ladder
computed with the DMRG and the RVA.

\begin{figure}
\vspace{-0.3cm}
\hspace{-0.5cm}
\epsfxsize=8cm \epsffile{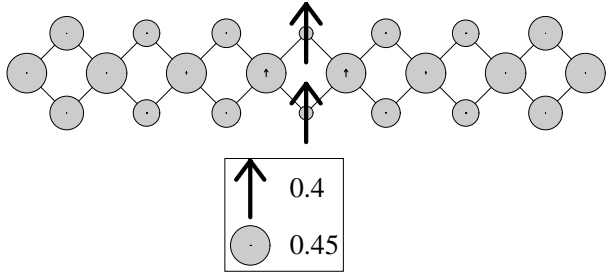}
\vspace{0.5cm}
\narrowtext
\caption[]{Results using DMRG showing the necklace state with 8 holes
and spin $S_z=1$. The diameter of the circles are proportional
to the hole density, and the length of the arrows are
porportional to $\langle S_z \rangle$, accoring to the scale in
the box.  }
\label{fig9} 
\end{figure}
\noindent

\item {\bf Case (8,1):} (See Fig.9)
The spin excitation of the $x=1/3$ state
is given by a  spin 1 magnon 
strongly localized  
on an odd parity plaquette located at the center
of the ladder. The other plaquettes
remain even and spinless. The value of the spin gap is
given by $\Delta_s= 0.27$ (DMRG) and 0.32 (RVA).

\begin{figure}
\vspace{-0.3cm}
\hspace{-0.5cm}
\epsfxsize=8cm \epsffile{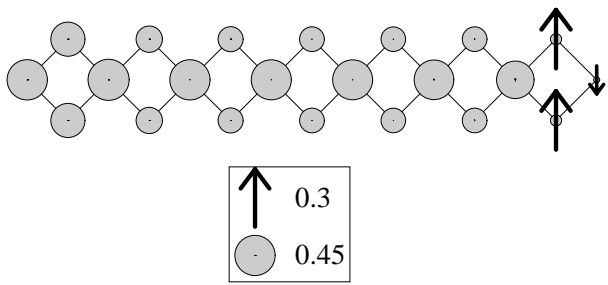}
\vspace{0.5cm}
\narrowtext
\caption[]{DMRG results for the hole and spin densities of 
the necklace state with 7 holes
and spin $S_z=1/2$.  }
\label{fig10} 
\end{figure}
\noindent

\item {\bf Case (7,1/2):}(See Fig.10)
This case is obtained by doping the $x=1/3$ state with an electron. 
The additional 
electron goes into either of the 
boundary plaquettes. The corresponding 
plaquette  changes its parity to $-1$.

\item {\bf Cases (6,0) and (6,1)}: (See Figs. 11-12)
The state $x=1/3$ is now doped with two electrons, which 
go to the boundary plaquettes which change their parity. 
There seems to be a
small effective coupling between the two spin 1/2 at the ends
of the ladder, which lead to a breaking of the degeneracy between the triplet
and the singlet. This is reminiscent  
of the effective spin 1/2 at 
the ends of the Haldane and AKLT open spin chains \cite{AKLT}. 
There also exists a weak effective coupling that breaks the 
four fold degeneracy of the open chains.

\begin{figure}
\vspace{-0.3cm}
\hspace{-0.5cm}
\epsfxsize=8cm \epsffile{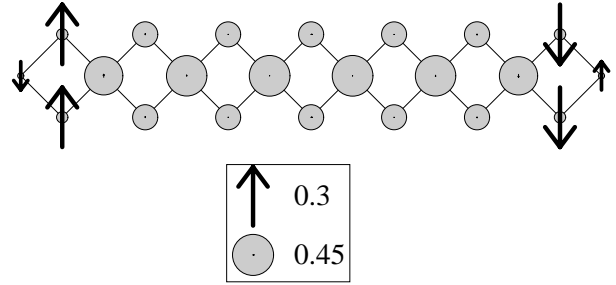}
\vspace{0.5cm}
\narrowtext
\caption[]{Results from DMRG showing the necklace state with 6 holes
and spin $S_z=0$. 
}
\label{fig11} 
\end{figure}
\noindent

\begin{figure}
\vspace{-0.3cm}
\hspace{-0.5cm}
\epsfxsize=8cm \epsffile{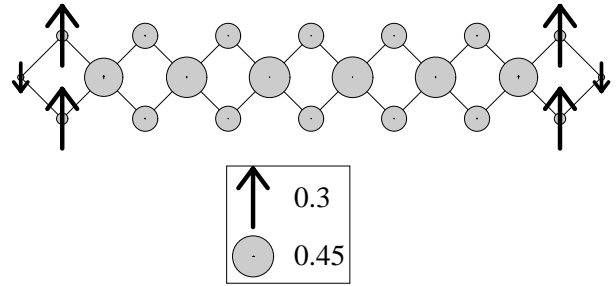}
\vspace{0.5cm}
\narrowtext
\caption[]{Results from DMRG showing the necklace state with 6 holes
and spin $S_z=1$. 
}
\label{fig12} 
\end{figure}
\noindent

\item {\bf Case (9,1/2):} (See Fig.13)
The $x=1/3$ state is doped with one hole. 
The parity of the plaquettes remain unchanged  
and the extra spin 1/2 delocalizes along the whole 
system perhaps with some SDW component.
The differences between this hole doped case 
and the electron doped case (7,1/2) are quite striking.

\item {\bf Case (10,0)} (See Fig.14)
This state looks very much the same as the $x=1/3$ 
but just with more holes.

\item {\bf Case (10,1)} (See Fig.15)
Same pattern as 
in the (9,1/2) case with the spin delocalized over the whole
system.

\end{itemize}

In summary the DMRG results 
clearly suggest the existence of two distinct
regimes corresponding to dopings
$0 \leq x < 1/3$ and $ x \geq 1/3$. In the overdoped
regime the plaquettes are always even while 
in the underdoped
regime they can be even or odd. 
Phase separation into even and odd plaquettes may
also be possible. Our  results
at this moment are ambiguous and 
further numerical work is required.
The most important result is the peculiar structure
of the $x=1/3$ state, which we shall  study further
in the next two sections.

\begin{figure}
\vspace{-0.3cm}
\hspace{-0.5cm}
\epsfxsize=8cm \epsffile{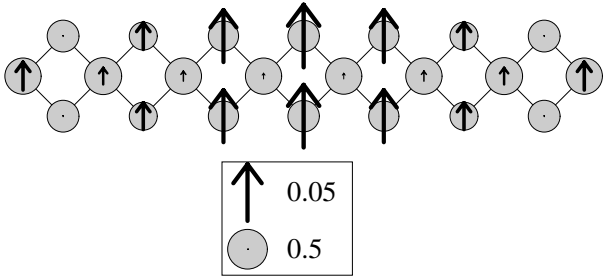}
\vspace{0.5cm}
\narrowtext
\caption[]{Results from DMRG showing the necklace state with 9 holes
and spin $S_z=1/2$. 
}
\label{fig13} 
\end{figure}
\noindent

\begin{figure}
\vspace{-0.3cm}
\hspace{-0.5cm}
\epsfxsize=8cm \epsffile{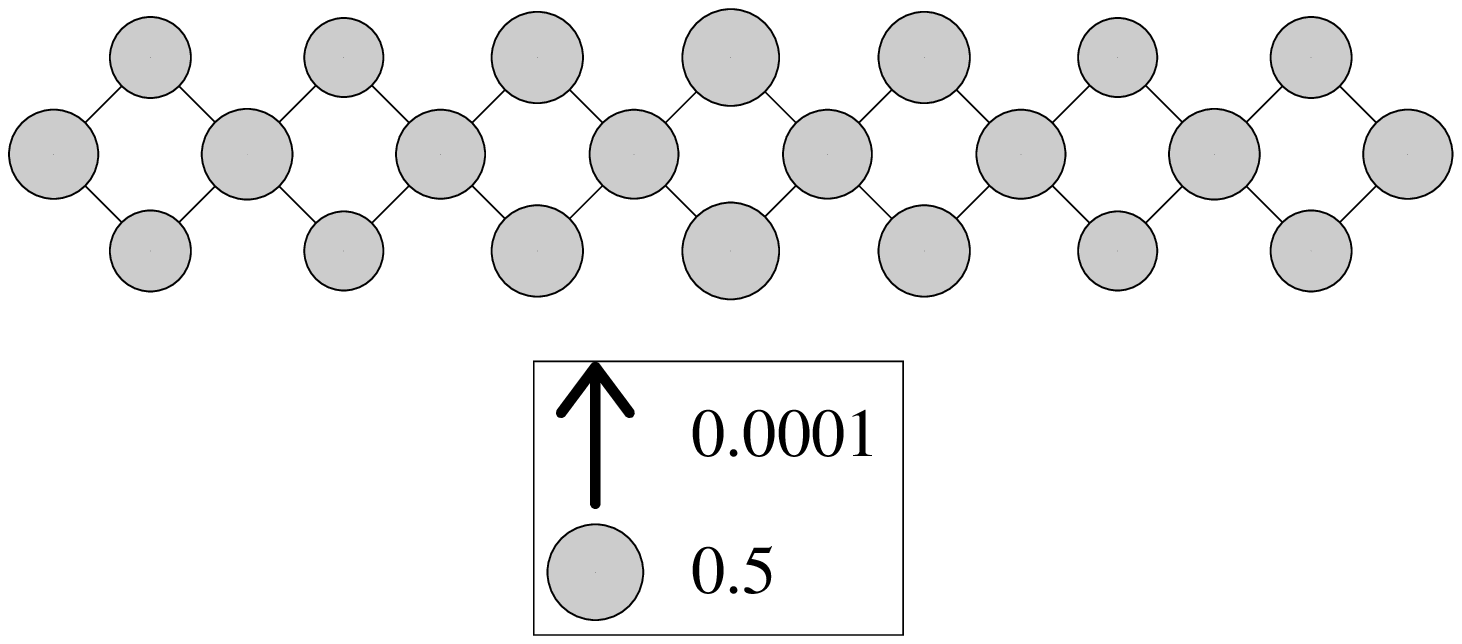}
\vspace{0.5cm}
\narrowtext
\caption[]{Results from DMRG showing the necklace state with 10 holes
and spin $S_z=0$. 
}
\label{fig14} 
\end{figure}
\noindent
\begin{figure}
\vspace{-0.3cm}
\hspace{-0.5cm}
\epsfxsize=8cm \epsffile{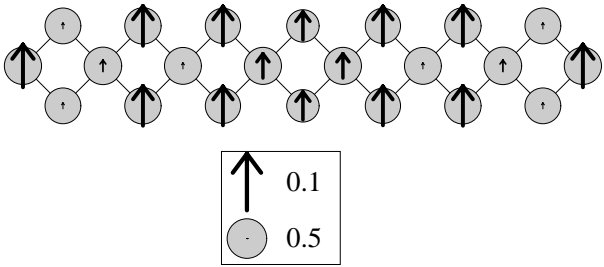}
\vspace{0.5cm}
\narrowtext
\caption[]{Results from DMRG showing the necklace state with 10 holes
and spin $S_z=1$. 
}
\label{fig15} 
\end{figure}
\noindent

\section{The $\lowercase{x}=1/3$ state of the necklace ladder}

The most important configuration contained in the
$x=1/3$ state has spin singlets along the diagonal. 
This is consistent with and helps explain the $\pi$ phase shift 
in the (1,1) domain walls observed numerically with  
DMRG\cite{WS-st} and Hartree-Fock\cite{hfdomain}  calculations in 
large lattices  and experimentally in
some nickelates compounds.

On the other hand
the $x=1/3$ state  is a kind of 1D generalization
of the GS of 2 holes and 2 electrons 
on  the $2 \times 2$ cluster discussed
in reference \cite{WS-lad}, in connection with the binding of holes
in  the 2-leg and higher-leg ladders. One can also
use this local structure to 
build up a variational 
state of the 2-leg ladder valid for any doping\cite{RVA2}.

\noindent

The GS of two holes in a plaquette 
is the localized Cooper pair 
depicted in Fig. 16
and can be generated by the pair field operators acting on the vacuum as,

\begin{eqnarray}
& | {\rm Cooper} \;{\rm Pair} > = & \label{11} \\
& \left( A ( \Delta_{1,3}^\dagger +
\Delta_{1,4}^\dagger + \Delta_{2,3}^\dagger +  \Delta_{2,4}^\dagger) 
+  (\Delta_{1,2}^\dagger + \Delta_{3,4}^\dagger) \right) |0\rangle &
\nonumber 
\end{eqnarray}

\noindent
where  $A$ is given by

\begin{equation}
A = \frac{1}{ [2+(J/4 t)^2 ]^{1/2} - J/4t}
\label{12}
\end{equation}

\begin{figure}
\hspace{-0.5cm}
\epsfxsize=9cm \epsffile{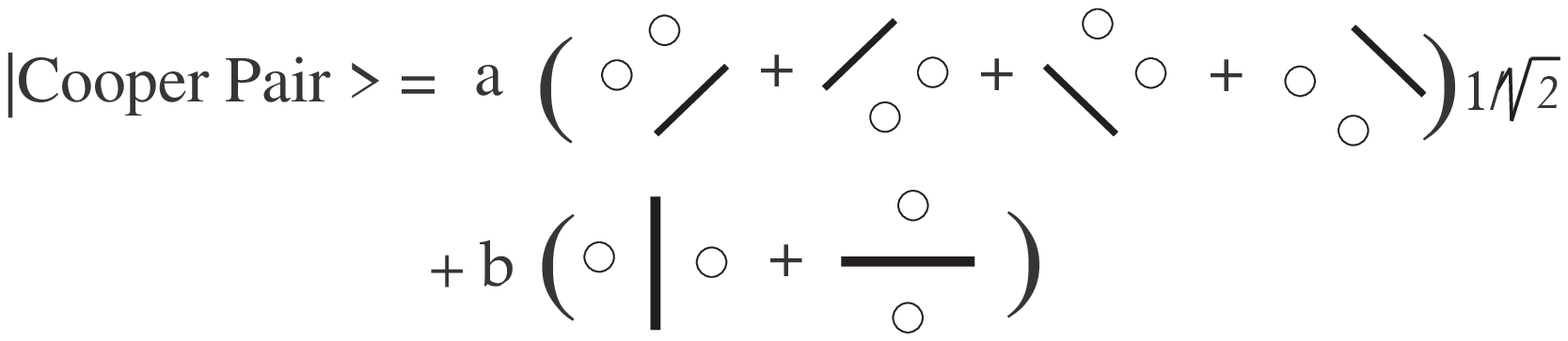}
\narrowtext
\caption[]{A pictorical representation of a localized Cooper pair on a plaquette.
The parameter $A$ in (\ref{11}) is $A= a/\sqrt{2}, b=1$.}
\label{fig16} 
\end{figure}
\noindent

\noindent
If $J/t<2$, then $ A < 1$ which  means that a diagonal bond
is more probable that a non-diagonal one.
This feature is also observed in the $x=1/3$ state where
the most probable 
bonds are those that line up along the transverse
diagonals of the ladder ( Fig. 7).
Taking into account this state together with some important local
fluctuations around it lead us to propose an 
RVA state with $x=1/3$, which is 
defined by the 
recursion relations depicted in Fig. 17 (see Appendix A for details).
The RVA state corresponding to $N=1$ plaquette coincides
precisely with state (\ref{11}) identifying $A= a/\sqrt{2}, b=1$ (See Fig.16).
For ladders with more than one plaquette ($N>1$)  
the symmetry of the diagonals
disappear and $b \neq 1$ in 
general. In this case we have two independent variational
parameters $a$ and $b$. 
In Fig. 18 we give
the energy per plaquette and the values of $a$ and $b$ as functions
of the number of plaquettes of the necklace ladder obtained
by minimization of the energy of the RVA state. We observe
that both $a$ and $b$ become less than one in agreement
with the DMRG results.
All these results are quite satisfactory but still they
do not give us a transparent physical picture of the $x=1/3$ state.
This will be done in the next section.

\begin{figure}
\hspace{-0.5cm}
\epsfxsize=9cm \epsffile{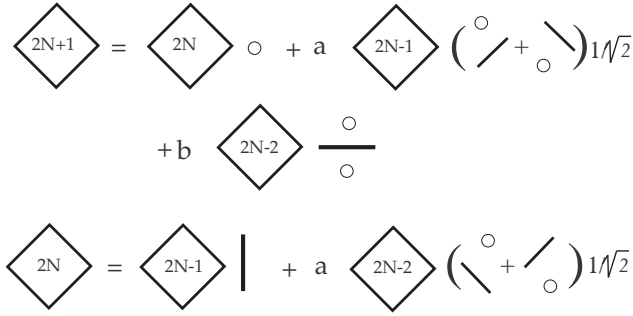}
\narrowtext
\caption[]{A pictorical representation of the Recurrence Relations employed
with the RVA method (See Appendix A) to construct variational G.S. states 
for the doped $x=1/3$ necklace ladder. The diagonal squares represent bulk states
of a given length.
Small circles represent holes and solid lines valence bonds.}
\label{fig17} 
\end{figure}
\noindent

\begin{figure}
\hspace{-1cm}
\epsfysize=9cm 
\epsfxsize=9cm \epsffile{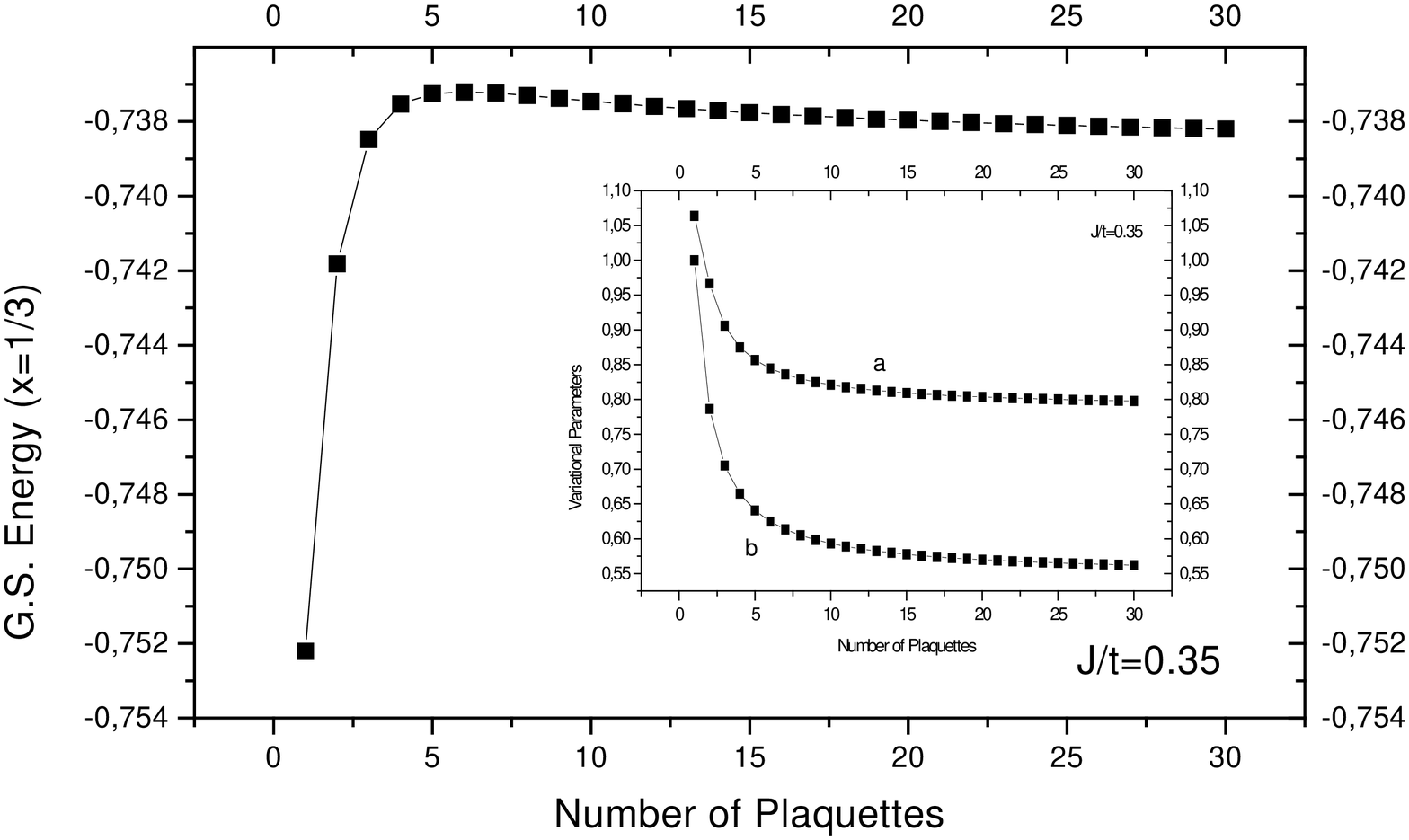}
\narrowtext
\caption[]{Ground state energy per site as function
of the number of plaquettes for a doped $x=1/3$
necklace ladder. It is obtained using
the RVA method (See Appendix A). Here $a$ and $b$ denote the 
variational parameters and it is shown how they get stabilized 
towards their thermodynamic values as the length of the ladder 
increases.}
\label{fig18} 
\end{figure}
\noindent

\section{The plaquette picture of the necklace ladder:
the $\lowercase{t_p}-J_{\lowercase{p}}-\lowercase{t_d}$ model}

An interesting property  of the rectangular ladders is  
that the  strong coupling picture 
of the GS and excited
states is generally valid also 
in the intermediate and weak coupling regimes.
Thus, for example, the spin gap of the 2-leg spin ladder
can be seen in the strong coupling limit 
as the energy cost for
breaking a bond along the rungs.

In Section II we suggested that diagonal ladders
could be thought of as collections of coupled 
plaquettes (Fig. 3).
The trouble is that in doing so one actually 
needs more sites than those
available in the original lattice. Indeed
the necklace ladder ${\cal L}^D$ with $N$ plaquettes
has $ 3N$ sites for $N$ large  while the extended or decorated ladder
${\cal L}^P$ shown on the right hand side of Fig. 3 
has $4N$.

The solution of this problem is achieved on 
physical grounds by defining  
on the lattice ${\cal L}^P$ an extended $t-J$
Hamiltonian  which, in a certain strong coupling  limit, becomes
equivalent to the standard $t-J$ Hamiltonian 
on ${\cal L}^D$. The
extended Hamiltonian can also be studied in the limit where 
the plaquettes are weakly coupled. 
As we shall see,
the latter
limit provides a useful physical picture 
 of the properties of the necklace ladder  
for $x=1/3$ and other dopings as well.

\subsection{The  $t_p-J_p-t_d$ model}

We shall define on the lattice ${\cal L}^P$ an extended
$t-J$ model by the following Hamiltonian,

\begin{eqnarray}
{H}_{pd}= &  H_p + H_{d} & \label{13} \\
H_p = & \sum_n \,\, h_n(t_p,J_p) & \nonumber \\
H_{d} = & \sum_{n} \,\, h_{n,n+1}(t_d)  \nonumber 
\end{eqnarray}

\noindent
where $h_n(t_p,J_p)$ 
is a standard $t_p-J_p$ Hamiltonian involving
only the 4 sites of the plaquette labelled by $n$. Of course
$h_n$ and $h_m$ commute for $n \neq m$. On the other
hand, $h_{n,n+1}(t_d)$ is a hopping Hamiltonian associated with 
the link 
that connects the two nearest neighbor
plaquettes $n$ and $n+1$. Denoting by $L$ and $R$ the corresponding
sites 
on the different plaquettes 
joined  by the link $<L,R>$ then $h_{n,n+1}$ is 
given by the link Hamiltonian defined as

\begin{equation}
h_{<L,R>}= t_d ( {c}^\dagger_{L,s} -  {c}_{R,s}^\dagger )  
\left(  {c}_{L,s} -  {c}_{R,s} \right)
\label{14}
\end{equation}

\subsection{Strong hopping limit of the  $t_p-J_p-t_d$ model}

We want to prove that in the strong hopping limit, where 
$t_d \rightarrow \infty$, the  
$t_p-J_p-t_d$ model becomes equivalent
to the $t-J$ model on the necklace ladder ${\cal L}^D$.

In this limit we first diagonalize $H_{d}$ looking for
the low energy modes of the plaquettes.
We then define  a renormalization group (RG) 
operator $T$, that leads to a renormalization
of operators in the extended lattice 
model into operators that act on the necklace lattice.
 In particular the $H_{pd}$ 
Hamiltonian is truncated to an effective Hamiltonian
which is equivalent to the original necklace 
Hamiltonian. The truncation operation is given by the Eq.
(for a review of the Real Space RG method see \cite{springer})

\begin{equation}
H_{\rm eff} = T \, {H}_{pd} \, T^\dagger
\label{15}
\end{equation}

The Hamiltonian $h_{n,n+1}(t_d)$ acts in a Hilbert space of
dimension $3 \times 3 = 9$. It has  two eigenvalues $E= 0$ and $2 t_d$,
with degeneracies 3 and 6 respectively. 
The zero eigenvalue
corresponds to the states with two holes and the bonding state 
with up and down spins. In the limit 
$t_d >> t_p, J_p$ one retains
only the latter  3 degrees of freedom which  can be thought of as  
 renormalized hole and spin up and spin down
electron states, respectively. The truncation operator
$T$, that maps the Hilbert space  
${\cal H}_{{\cal L}^P}$
into the effective Hilbert space ${\cal H}_{{\cal L}^D}$ is,
given by

\begin{equation}
\begin{array}{rcl} 
T :{\cal H}_{{\cal L}^P} & \rightarrow & {\cal H}_{{\cal L}^D} \\
|\bullet, \bullet\rangle & \rightarrow & 0 \\
|\bullet, \circ \rangle & \rightarrow & \frac{1}{\sqrt{2}} 
|\ast \rangle \\
|\circ, \bullet \rangle & \rightarrow & \frac{1}{\sqrt{2}}
|\ast \rangle \\
|\circ, \circ \rangle & \rightarrow & |o \rangle \end{array}
\label{16}
\end{equation}

\noindent where $\bullet$ and $\circ$ 
stand for one electron, with spin up or down,  
and one hole respectively living on a given link of
${\cal L}^P$, while 
$\ast$ and $o$ are the effective electron and holes living on the
corresponding site of ${\cal L}^D$ 
obtained by contracting the previous link to a site.
The hermitean operator $T^\dagger$ acts
as follows,

\begin{equation}
\begin{array}{rcl} 
T^\dagger :{\cal H}_{{\cal L}^D} & 
{\rightarrow} & {\cal H}_{{\cal L}^P} \\
|\ast\rangle & \rightarrow & \frac{1}{\sqrt{2}} 
\left( |\bullet, \circ \rangle + |\circ, \bullet \rangle \right)  \\
|o \rangle & \rightarrow & |\circ, \circ \rangle  \end{array}
\label{17}
\end{equation}

Eq.(\ref{17}) means that 
an electron  state $|\ast\rangle$ of ${\cal L}^D$ becomes the bonding
state in the enlarged Hilbert space ${\cal H}_{{\cal L}^P}$.
The RG operators $T$ and $T^{\dagger}$ 
defined above satisfy the following Eqs.\cite{springer}

\begin{equation}
T \; T^\dagger = {\bf 1}, \;\;\;\; T^\dagger \; T = P_{G}^{(\ell)}
\label{18}
\end{equation}

\noindent
where $P_{G}^{(\ell)}$ is a Gutzwiller operator which now acts on links
rather than on sites as follows,

\begin{equation}
\begin{array}{rcl} 
P^{(\ell)}_G: {\cal H}_{{\cal L}^P} 
& \rightarrow & {\cal H}_{{\cal L}^P} \\
|\bullet, \bullet\rangle & \rightarrow & 0 \\
|\bullet, \circ \rangle & \rightarrow & \frac{1}{\sqrt{2}} 
\left( |\bullet, \circ \rangle + |\circ, \bullet \rangle \right)  \\
|\circ, \bullet \rangle & \rightarrow & \frac{1}{\sqrt{2}} 
\left( |\bullet, \circ \rangle + |\circ, \bullet \rangle \right)  \\
|\circ, \circ \rangle & \rightarrow & |\circ, \circ \rangle \end{array}
\label{19}
\end{equation}

Using the above definitions we can easily 
obtain the renormalization
of the different operators acting in 
${\cal L}^P$,

\begin{equation}
\begin{array}{rl} 
T \; {c}_{L,s} \; T^\dagger =
T \; {c}_{R,s} \; T^\dagger =
 & \frac{1}{\sqrt{2}} c_{M,s} \\
T \; {{\bf S}}_L \; T^\dagger = 
T \; {{\bf S}}_R \; T^\dagger = 
& \frac{1}{2} {\bf S}_{M} \\
T \; {n}_L \; T^\dagger = 
T \; {n}_R \; T^\dagger = 
& \frac{1}{2} n_{M}\ . \end{array}
\label{20}
\end{equation}

\noindent Here ${c}_{i,s}, {{\bf S}}_i, {n}_i$ are
the fermion, spin and number operators acting at the
edges of the link $<L,R>$ for $i=L,R$ while for $i=M$
they act at the effective ``middle" point of the link
(i.e. $<L,R> \rightarrow M$).
Of course, the operators and states
that are not 
on the principal diagonal of both ${\cal L}^P$ and ${\cal L}^D$
are not affected by  the renormalization procedure.

Using Eqs (\ref{15}) and (\ref{20})  
we can immediately find 
that the renormalized effective
Hamiltonian is  given by 
the $t-J$ Hamiltonian (\ref{4}), i.e.

\begin{equation}
H_{\rm eff} \equiv  T \; {H}_{pd} \; T^\dagger
= T \; H_p ( t_p,J_p) \; T^\dagger
= H_{t,J}
\label{21}
\end{equation}

\noindent
with the following 
values for the coupling constants,

\begin{equation}
t= \frac{1}{ \sqrt{2}} \; t_p ,\;\;\; \;\; 
J = \frac{1}{ {2}} \; J_p, \;\;\; ({\rm necklace}\;\;{\rm ladder})
\label{22} 
\end{equation}

\noindent In the derivation of (\ref{22}) 
we are assuming periodic boundary conditions,
along the principal diagonal of the necklace ladder.

The strong hopping limit studied above
is reminiscent of the  
strong coupling  limit of 
 the Hubbard model which leads to the t-J model
plus some extra three-site  terms which are  usually ignored. 
In the latter case the strong
Coulomb  repulsion forces the Gutzwiller on-site
constraint.
Our case is a ``dual version" of this mechanism, in the sense
that the coupling constant involved is a hopping 
parameter, and
that the Gutzwiller constraint arises from a link rather
than from a site constraint. In the case of the $t_p-J_p-t_d$ 
model one does
not have to do perturbation theory in order 
to produce the exchange term in the effective
Hamiltonian since it is already contained in the
plaquette Hamiltonian. 
Perturbation theory would  
produce terms of the order $1/t_d$, but they vanish
at $t_d=\infty$.  The construction we have performed in this
section can in principle be generalized to the Hubbard model\cite{hubbard}.

The analogy between  the Hubbard and the $t_p-J_p-t_d$ 
model suggests that
we may learn something about the  strong hopping limit 
by studying the weak hopping one.
This is certainly true if there are no phase transitions between 
the two regimes.

\subsection{Weak Hopping Limit of the $t_p-J_p-t_d$ model}

In the weak hopping regime, i.e. $t_d << t_p, J_p$, 
we first diagonalize the plaquette Hamiltonian $H_p$
and treat $H_d$ as a perturbation.
The energy levels of $H_p$ 
are given, to lowest order in perturbation theory,
as tensor products of the eigenstates of every
plaquette. 
There will be in general a huge degeneracy, 
which will be
broken by the effective Hamiltonian derived from $H_{d}$ using
perturbation theory.
Before going further into the study of the plaquette Hamiltonian
we  have to consider the relationship between
the filling  factors of  the states 
belonging to lattices with different number of sites.

Let us consider a state in ${\cal L}^D$ 
with $N_h$ holes and
$N_e$ electrons. Applying the operator $T^\dagger$, 
this state is transformed to a state 
in   ${\cal L}^P$ with $N^{(p)}_h$ holes and $N^{(p)}_e$
electrons given by,

\begin{equation}
N_h^{(p)} = \frac{4}{3}  N_h + \frac{1}{3} N_e, 
\;\;{N_e}^{(p)} = N_e
\label{23}
\end{equation}
These equations reflect  the fact 
one gets an extra hole upon  going to the enlarged lattice.
Eqs.(\ref{23}) imply the following relations between
the doping factors $x=N_h/(N_h + N_e)$ and  ${x}_p
 ={N_h}^{(p)}/({N_h}^{(p)} + {N_e}^{(p)})$,

\begin{equation}
\;x = 
\frac{1}{3} ( 4 {x}_p - 1), \;\;\; 
{x}_p =\frac{1}{4} ( 1 + 3 \;x) \;\; \;
\label{24}
\end{equation}

\noindent
From (\ref{24}) we get the following correspondences

\begin{eqnarray}
& x= 0 \longleftrightarrow {x}_p = 1/4 & \label{25} \\
& x= 1/3 \longleftrightarrow {x}_p = 1/2 & \nonumber 
\end{eqnarray}

\noindent which we shall discuss in detail below.

{\bf Weak hopping picture of the $x=1/3$ state}

Eq.(\ref{25}) implies that the 1/3 doped state 
of the necklace ladder is transformed to a state
with two holes and two electrons 
per plaquette in the expanded necklace lattice. 
We show in appendix B that the lowest GS for this filling
is given by the coherent superposition of  Cooper pairs 
localized on the plaquettes, i.e.

\begin{equation}
|{x}_p=1/2, t_d =0> = \prod_{n=1}^N \;\; |{\rm Cooper} \;{\rm pair}>_n
\label{26}
\end{equation}

\noindent
where 
$ |{\rm Cooper} \;{\rm pair}>_n$ is the state given in Eq..(\ref{11})
with the parameters $a$ and $b$ given by Eq.(\ref{12}) for the
values of $t_p, J_p$. 

\begin{figure}
\hspace{-1cm}
\epsfxsize=9cm \epsffile{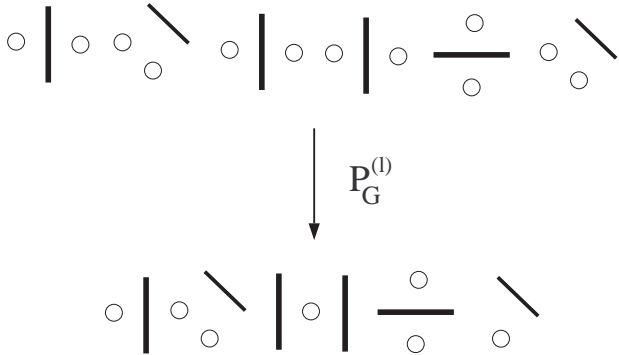}
\narrowtext
\caption[]{Ground state for a doped $x=1/3$
necklace ladder (down) obtained as the projection of 
a $x_p=1/2$ doped state in the decorated (dual) diagonal ladder (above).}
\label{fig19} 
\end{figure}
\noindent

Turning $t_d$ on, the state (\ref{26}) will be perturbed
mainly along the principal diagonal. The doubly occupied and antibonding
links will become high energy states while
the bonding and empty links will remain low in energy.
In the limit when $t_d$ becomes infinite we expect 
the GS (\ref{26}) to evolve continuously into the $x=1/3$ 
GS of the necklace. This suggest that the $x=1/3$ state
of the necklace ladder can  be described as  a Gutzwiller 
projected state, i.e.

\begin{equation}
|x=1/3> \sim T\; {\cal P}_G^{(\ell)} \; |{x}_p=1/2,t_d=0>
\label{27}
\end{equation}

\noindent where  we first project out the doubly occupied 
and antibonding states on the links on the principal
diagonal of the expanded ladder and then project
the resulting state into the Hilbert space of the necklace
ladder.
Some diagrammatics (Fig. 19) shows  that 
the state (\ref{27}) is basically the same
as the $x=1/3$ 
RVA state constructed  in Section VI.
This leads us to  the conclusion that the $x=1/3$ GS
of the necklace ladder  
can be seen as  the Gutzwiller projection  
of Cooper pairs
localized on the plaquettes. 
In this case the
Cooper pairs are locked in a Mott insulating phase 
and there is an 
exponential decay of the pair field.

{\bf Weak hopping picture of the $x=0$ state}

The GS of a plaquette with 
1 hole and 3 electrons for $J_p/t_p=0.5$ 
has spin 1/2 and it belongs to the 2-dimensional  irrep labelled 
by $E$ of the symmetry
group $D_4$ (see Appendix B). 
These two states differ in their parity  
along the minor diagonal which can be even or odd.
Both  states can be thought of as  bound states of a
Cooper pair and one electron ( see Fig. 20).
The four fold degeneracy on every plaquette
is broken by $t_d$. The odd parity plaquettes are lower in energy
than the even ones and the effective model is given by a
ferromagnetic spin 1/2 chain,

\begin{equation}
H_{\rm eff} = J_{\rm eff} \;\; \sum_n \; {\bf S}_{n}^{\rm eff} 
\cdot {\bf S}_{n+1}^{\rm eff} 
\label{28}
\end{equation}

\begin{figure}
\hspace{-1cm}
\epsfxsize=9cm \epsffile{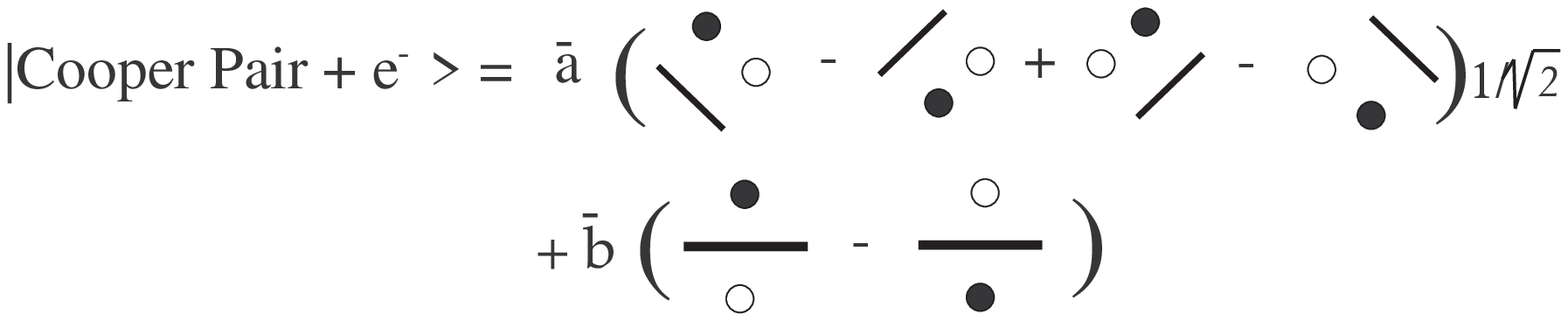}
\narrowtext
\caption[]{A pictorical representation of a bound state formed
by a Cooper pair and one electron. Small blank circles represent
holes, black circles represent electrons and solid lines are
valence bonds. Here $\bar{a}$, $\bar{b}$ are relative amplitudes
taken as variational parameters.}
\label{fig20} 
\end{figure}
\noindent

\noindent
Here $ {\bf S}_{n}^{\rm eff} $ is the overall spin 1/2 
operator of the odd plaquette and $J_{\rm eff} \sim - t_d^2$
is a ferromagnetic exchange coupling constant.

Following a reasoning  similar to that for the case $x=1/3$ 
we conjecture that the $x=0$ state 
can be represented as the following 
Gutzwiller projected state,

\begin{equation}
|x=0> \sim T\; {\cal P}_G^{(\ell)} \; 
|{x}_p = 1/4,\epsilon_n=-1, t_d=0>
\label{29}
\end{equation}

\noindent whose structure is indeed very similar 
to the ferrimagnetic RVA state
proposed in Section III. See Fig. 21 for a plaquette 
construction
of the N{\'e}el state of the necklace ladder starting from
the ${x}_p=1/4, \;\; \epsilon_n=-1$ state. 
The gapless excitations
of the ferrimagnetic $x=0$ GS correspond, in the weak coupling
picture, to the magnons of the ferromagnetic chain (\ref{29}),
while the gapped excitations correspond to an excitation
of the plaquette to a state with spin 3/2. 

\begin{figure}
\hspace{-1cm}
\epsfxsize=9cm \epsffile{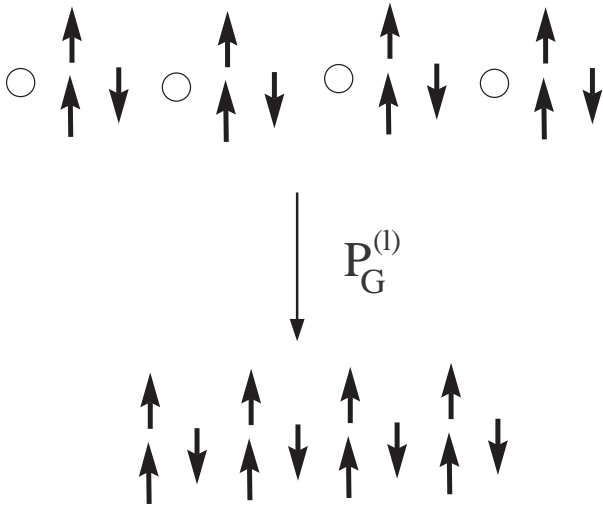}
\narrowtext
\caption[]{A Neel-like ground state (down) for an undoped $x=0$ 
necklace ladder obtained as the projection (above) of a doped state
in the decorated (dual) diagonal ladder.}
\label{fig21} 
\end{figure}
\noindent

In summary we have been able to obtain a satisfactory
picture of both the $x=0$ and 
1/3 states in the weak coupling limit
of the extended $t-J$ model, which leads us to conclude 
that for these dopings 
there are no phase transition between the weak
and strong coupling regimes.
Other dopings 
involve the competition of the two elementary plaquettes states
used above
and will be considered elsewhere.

\section{From 1D to 2D through diagonal ladders}

The necklace ladder represents the first step in the diagonal route
to the 2D square lattice. In this section we shall push forward
this viewpoint trying to see how much one can expect from it. 
This will lead us to ask questions whose solution we do not
yet know. In this sense 
some of the material presented below is conjectural.

Let us first start with a
short  excursion into  graph theory.

\subsection{The plaquette construction and medial graphs}

The plaquette construction of the necklace ladder
is related to the so called medial graphs
used in the coloring problem or in Statistical
Mechanics \cite{Bax}. Before we show this connection
we need to generalize our plaquette construction to
diagonal ladders with more than one plaquette per unit cell
, i.e. $n_p > 1$.

In this section we shall use the following notations,

\begin{equation}
\begin{array}{l}
{\cal L}^R_{n_\ell}: \;\; {\rm rectangular} \;\;{\rm ladder} \;\;
{\rm with} \;\;n_\ell \;\;{\rm legs}\\
{\cal L}^D_{n_p}: \;\; {\rm diagonal} \;\;{\rm ladder} \;\;
{\rm with} \;\; n_p \;\;{\rm plaquettes} \\
{\cal L}^P_{n_p}: \;\; {\rm (4,8)} \;\;{\rm lattice} \;\;
{\rm with} \;\; n_p\;\;{\rm plaquettes} \\
\end{array}
\label{30}
\end{equation}

\noindent As an example we depict 
in Fig. 22 the lattices ${\cal L}^P_2, {\cal L}^D_3$
and ${\cal L}^R_2$.
The  lattice ${\cal L}^P_{n_p}$ consist of 4-gons, i.e. plaquettes,
joined by links, which are associated with the hopping
parameter $t_d$ while the plaquettes are associated with 
the parameters $t_p$ and $J_p$.
For $n_p >1$ ${\cal L}^P_{n_p}$ contains 
also 8-gons that are formed by
four $t_d$ links and four $t_p$ links. 

\begin{figure}
\hspace{-1cm}
\epsfxsize=9cm \epsffile{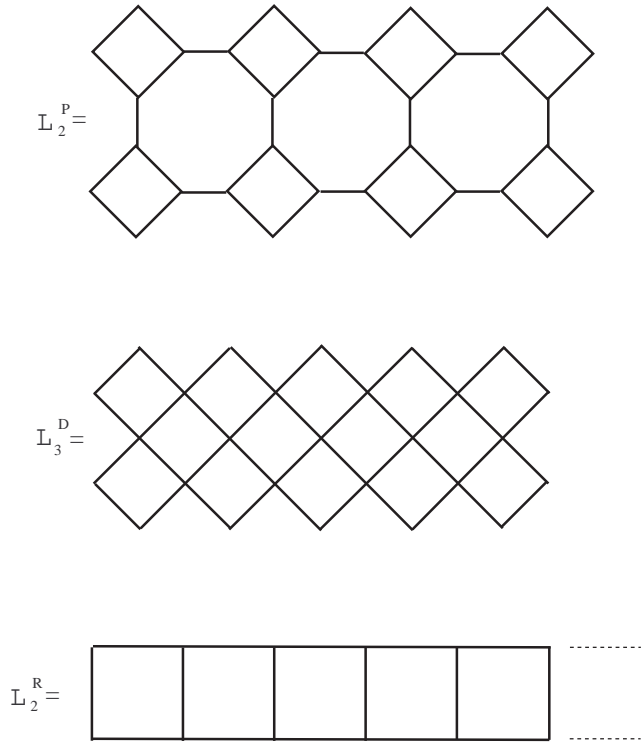}
\narrowtext
\caption[]{Several examples of related ladders as explained in text.
(Above) An example of $(4,8)$-lattice with $n_p=2$ plaquettes.
(Middle) An example of diagonal ladder with $n_p=3$ plaquettes.
(Down) An example of rectangular ladder with $n_=2$ legs.}
\label{fig22} 
\end{figure}
\noindent

As shown in the previous section the limit $t_d \rightarrow \infty$
has the geometric significance of shrinking the corresponding
$t_d$-links into sites, so that  the  
the lattice 
${\cal L}^P_{n_p}$ ``renormalizes" into  the diagonal ladder
${\cal L}_{2 n_p-1}^D$ (see Fig. 22). 
In this strong coupling limit the number 
of plaquettes actually
increases and  some plaquettes are generated for free.
The number of plaquettes of the diagonal ladder so
obtained is odd. This construction does not produce
even plaquette diagonal ladders.

Observe that all the
diagonal ladders are bipartite lattices but only when
$n_p$ is even are the number of sites of the two different
sublattices the same. This suggests that the $n_p$-even
diagonal AFH ladders belong to the same universality class as the
$n_l$-even rectangular ladders, while the $n_p$-odd diagonal ladders
belong to a different universality 
class characterized by ferrimagnetic GS's.

The opposite limit, where   $t_d \rightarrow 0$,  
has the geometrical meaning of 
shrinking the plaquettes into sites, so that 
${\cal L}^P_{n_p}$ 
``renormalizes'' the system to a rectangular  ladder
with $n_p$ legs ( see Fig. 22).

We summarize the above geometric RG operations  in the following
symbolic manner,

\begin{equation}
\begin{array}{lcll}
{\cal L}^P_{n_p} & \longrightarrow &
{\cal L}_{2 n_p -1}^D &  (t_d \rightarrow \infty) \\
{\cal L}^P_{n_p} & \longrightarrow &
{\cal L}_{n_p}^R &  (t_d \rightarrow 0) \end{array}
\label{31} 
\end{equation}

\noindent
In this sense  the $(4,8)$ lattice is an interpolating 
structure between diagonal and rectangular lattices.

There is an  interesting
connection between this
plaquette construction  
and the theory of medial graphs.
Consider a graph $G$  made of a set of
points $i$ connected by links $<i,j>$.
A medial graph ${\cal M}(G)$, associated with the graph
$G$, is obtained by surrounding every site $i$ of $G$ by
a polygon $P_i$,
such that two polygons $P_i$ and $P_j$, which
correspond to a link $<i,j>$, meet at a single  
intersection point $P_i \cap P_j$, which lies  on the middle
of the link $<i,j>$\cite{Bax} (see Fig. 23 for a generic example).

Choosing the polygons  $P_i$ to be 4-gons, i.e. plaquettes,  
one can easily show  that
a diagonal ladder with an odd number of
plaquettes is the medial graph of a rectangular
ladder, namely

\begin{equation}
{\cal L}^D_{2 n_p-1} = \;{\cal M}( {\cal L}^R_{n_p} )
\label{32}
\end{equation}

Medial graphs are used  in
Statistical Mechanics  to show the equivalence between the Potts
model and the 6-vertex model\cite{TL,Bax}. 
Indeed one can show that
the Potts model defined on a graph
$G$ is equivalent, i.e. has the same partition function
after appropriate identification of parameters, to
the 6 vertex model defined on the medial graph ${\cal M}(G)$, i.e.

\begin{equation}
Z_{\rm Potts}(G) = Z_{6-{\rm vertex}}( {\cal M}(G))
\label{33}
\end{equation}

\noindent
The transformation $G \rightarrow {\cal M}(G)$ 
is a kind of
duality map that relates two seemingly unrelated models
and it is in fact the key to solve the 
2D critical Potts model in terms of the 6 vertex one.

\begin{figure}
\epsfxsize=8cm \epsffile{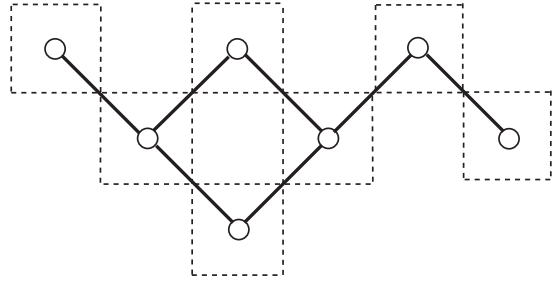}
\narrowtext
\caption[]{An example of medial graph construction.
Here the graph G is made by solid lines and blank circles.
Its associated medial graph ${\cal M}(G)$ is made by
dashed lines.}
\label{fig23} 
\end{figure}
\noindent

\subsection{Plaquette Construction of the 2D Square Lattice}

In Fig. 24 we apply the plaquette construction to the
2D square lattice. It is a simple generalization 
of the construction shown in the previous subsection 
when $n_p \rightarrow \infty$. If ${\cal L}^D_{\infty}$ is
a square lattice with lattice spacing $a$, then
${\cal L}^R_{\infty}$ is also a square lattice but with
spacing $\sqrt{2} a$ \cite{comment2}. 

\begin{figure}
\epsfxsize=8cm \epsffile{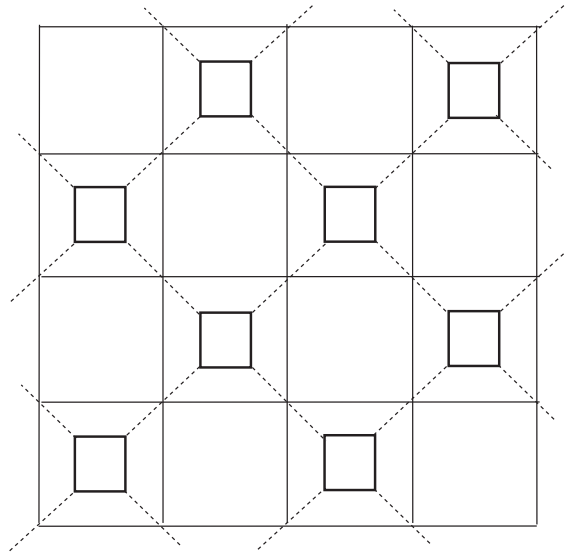}
\narrowtext
\caption[]{Plaquette constrution (small interior squares plus dashed lines)
 of the 2D square lattice as explained in text.}
\label{fig24} 
\end{figure}
\noindent

Let $x$ be the doping of a $t-J$ model defined
on  ${\cal L}^D_{\infty}$, and ${x}_p$ the
doping factor of a $t_p-J_p-t_d$ 
model defined on ${\cal L}^P_\infty$, then
the relations between these quantities are
analogous to Eqs.(\ref{22}) and (\ref{24}) for the necklace
ladder, namely

\begin{equation}
x =  (2 {x}_p -1) 
\label{34}
\end{equation}

\begin{equation}
t = \frac{1}{2} \; t_p,\;\;\;
J = \frac{1}{4} \; J_p 
\label{35}
\end{equation}

Eq.(\ref{34}) implies that the undoped system $x=0$ corresponds
to doping $x_p=1/2$ in the enlarged lattice.
Fig. 25 shows a plaquette 
construction of the    N{\'e}el state from the $x_p=1/2$ state.
Notice that the plaquettes have spin 1 and that the parity
on their diagonals alternate between $(1,-1)$ and $(-1,1)$.
In the strong hopping limit  
the plaquettes 
have an effective 
spin 1. The whole set of these effective spin 1's
are coupled antiferromagnetically and form a
square lattice with lattice spacing which is  $\sqrt{2}$ times
larger than the lattice spacing of the original spin 1/2 model. 
In a certain sense the plaquette construction 
integrates out degrees of freedom and renormalizes the system into
an AF Heisenberg model with spin 1 and lattice space $\sqrt{2} a$.
This picture agrees qualitatively with the RG flow of the
$O(3)$ non linear sigma model in the renormalized classical
region at zero temperature\cite{CHN}. 

\begin{figure}
\hspace{-1cm}
\epsfxsize=9cm \epsffile{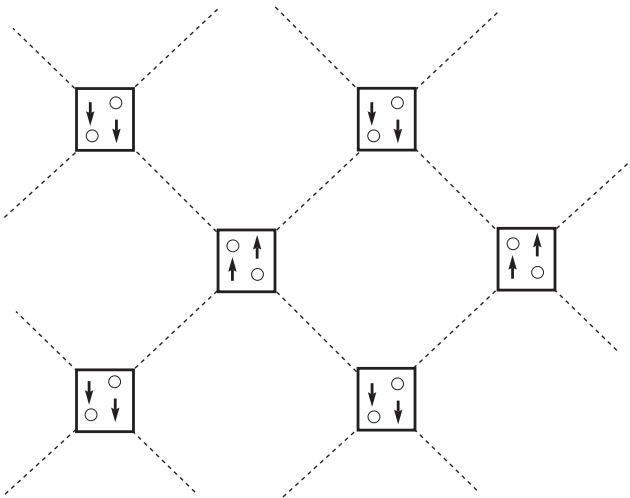}
\narrowtext
\caption[]{Plaquette constrution 
 of the 2D Neel state in a square lattice as explained in text.
The corresponding depicted state on the decorated (dual) lattice
has doping $x_p=1/2$ and spin 1 on every small square plaquette.}
\label{fig25} 
\end{figure}
\noindent

In the weak hopping limit, 
however the GS of the 2D model 
is given essentially by the coherent superposition of 
localized Cooper pairs
used in the construction of the $x=1/3$ necklace state. 
The Gutzwiller projection of this state onto the 
original lattice will produce Spin Peierls 
state rather than an AFLRO state.  

We conclude that unlike  
the case of the necklace ladder, 
the $t_p-J_p-t_d$ model in 2D
must have a phase transition
for some intermediate value of $t_d$. The study of this model
may serve to clarify the relationship between the AFLRO and 
the d-wave pairing structures observed in the theoretical 
models of strongly correlated systems.

\section{Conclusions}

Diagonal $t-J$ ladders provide an alternative route of interpolating between
1 and 2 spatial dimensions. Here we have described a general framework for
such an interpolation and introduced a generalized $t_p-J_p-t_d$ plaquette
model in which the individual plaquettes are linked by a hopping term
$t_d$. In the strong 
hopping $t_d \gg t_p, J_p$ limit, the generalized plaquette
model was shown to map into the original diagonal $t-J$ model with
renormalized parameters and filling factor. Thus, the generalized
$t_p-J_p-t_d$ model provides a dual model to the original diagonal model.
In this sense, it is interesting to study the $t_p-J_p-t_d$ model in the
weak hopping limit.  
If there is no phase transition between the weak and
strong hopping limits, then the weak hopping limit can provide new
insight into the nature of the original $t-J$ diagonal ladder.  We believe
that this is the case for the $n_p=1$ plaquette necklace ladder and that
its groundstate for a doping $x= \frac{1}{3}$ can be understood as the
Gutzwiller projection of a product state of Cooper pairs localized on the
plaquettes of the quarter-filled extended $t_p-J_p-t_d$ model. Alternatively,
for the $n_p\to \infty$ 2D limit, we believe that the extended
$t_p-J_p-t_d$ model at a doping $x_p=\frac{1}{2}$, which corresponds to the
undoped $(x=0)$ $t-J$ model, will have a phase transition for an
intermediate value of $t_d/t_p$. In this case, our conjecture is that the
strong coupling limit will have a ground state with long range AF order
while the weak coupling phase will be a localized Spin Peierls state.

In order to make these ideas more concrete, we have focused on the single
plaquette $n_p=1$ necklace ladder.  Here, using the results of DMRG and RVA
calculations, we have studied the necklace ladder for various dopings $x$.
For $x=\frac{1}{3}$, the DMRG calculations show that in the most probable
configuration, the holes occupy the sites along the principal diagonal of
the necklace and the spins form perfect singlets along the minor diagonals.
The RVA calculations, starting from a ``classical''
configuration and mixing in local quantum fluctuations about this state,
provide a ground state energy in good agreement with the DMRG result. Then, as
discussed above, a more transparent physical picture of the $x=\frac{1}{3}$
state of the diagonal necklace is provided by the extended $t_p-J_p-t_d$
dual model at a filling of $x_p=\frac{1}{2}$ 
in which this state is seen as a localized Cooper pair state.
It will be interesting to understand what happens when
additional holes are added. In particular, will a necklace with
a doping of $x_p=0.5 + \delta$ have power law $d$-wave like
pairing correlations?

For $x=0$, the diagonal necklace is equivalent to an alternating
$s=1/s=\frac{1}{2}$ spin chain and has a ferrimagnetic ground state with total
spin $N/2$, with $N$ the number of unit cells of the necklace. There are
gapped excitations with spin $N/2+1$ and gapless excitations with spin
$N/2-1$. In the weak hopping limit of the $t_p-J_p-t_d$ model, these
excitations correspond to local excitations of the plaquettes to a spin
$3/2$ state and to magnons of a ferromagnetic spin $\frac{1}{2}$ chain
respectively. Thus, in the $x=0$ case, the dual model provides a
useful physical picture.

We also have found that when the $x=\frac{1}{3}$ state is doped with holes,
the ground state plaquettes retain the even parity characteristic of the
$x=\frac{1}{3}$ state. However, when electrons are added, this parity can
be even or odd. Thus, it appears that the $x=\frac{1}{3}$ doping separates
the system into two distinct regions.

Clearly, the diagonal ladders form a rich class of models with properties
ranging from ferrimagnetic to antiferromagnetic and from localized pair states
to possible extended pairing states. Furthermore, the
$t_p-J_p-t_d$ model
provides a dual description which suggests alternative physical pictures
and approximation schemes as well as connections to concepts from
statistical mechanics.

{\bf Acknowledgements} 
We would like to acknowledge useful discussions with Hsiu-Hau
Lin and Eric Jeckelmann. 
GS would like to thank the members of the Physics Department
of the UCSB for their warm hospitality. GS and MAMD acknowledges
support from the DGES
under contract PB96-0906, SRW acknowledges support
from the NSF under Grant No.DMR-9509945, DJS acknowledges
support from the NSF under Grant No. DMR-9527304
and JD acknowledges support from the DIGICYT under contract PB95/0123.

\section*{Appendix A: RVA approach to the necklace  
ladder}

The RVA method is a kind of simplified DMRG 
where one retains a single state as the best candidate
for the GS of the system. As in the DMRG the GS
of a given length is constructed recursively from the states
defined in previous steps. This idea can be implemented
analytically if the ansatz is sufficiently simple. 
Below we shall propose various RVA states for the necklace
ladder with dopings $0 \leq x \leq 1/3$.

\begin{figure}
\epsfxsize=9cm \epsffile{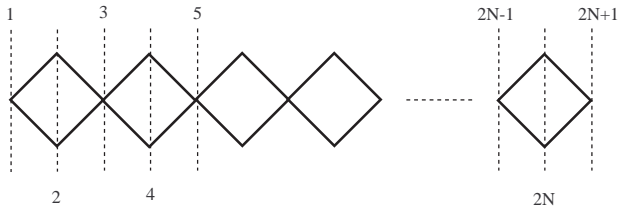}
\narrowtext
\caption[]{A pictorical view of the necklace ladder showing
the labelling convention employed to denote the variational 
RVA states. Here the positions along the main (horizontal) diagonal
of the ladder are sing odd sites, while the even positions are made
up of rungs or sudiagonals.}
\label{fig26} 
\end{figure}
\noindent

\subsection{ Case $x=0$}

Let us begin by labelling the sites of the necklace
ladder as in Fig. 26. The even sites denote the minor
diagonal of the ladder while the odd sites are those
on the principal diagonal.

At zero doping there are only two 
possible  states on the odd sites given by,

\begin{equation}
|\uparrow  > = c^\dagger_{\uparrow} \;|0>, \;\;\;\
|\downarrow  > = c^\dagger_{\downarrow} \;|0> 
\label{a1}
\end{equation}
On the even sites there are a triplet and a singlet state given by,

\begin{equation}
\begin{array}{cl}
|S >= & \Delta_{1,2}^\dagger \;|0> \\
|T_{\uparrow} > =& c^\dagger_{1,\uparrow} \;
c^\dagger_{2,\uparrow} \;|0> \\
|T_{\downarrow} > = & c^\dagger_{1,\downarrow} \;
c^\dagger_{2,\downarrow} \;|0> \\
|T_{0} > =& \frac{1}{\sqrt{2}} (c^\dagger_{1,\uparrow} \;
c^\dagger_{2,\downarrow} - 
c^\dagger_{1,\downarrow} \;
c^\dagger_{2,\uparrow}) 
\;|0> \end{array}
\label{a2} 
\end{equation}

The N{\'e}el state on the necklace ladder can be written 
simply as,

\begin{equation}
|\downarrow> \; |T_{\uparrow}> \; 
|\downarrow> \; |T_{\uparrow}> \; \cdots
 |\downarrow> \; |T_{\uparrow}> \;
 |\downarrow> \; |T_{\uparrow}> \
\label{a3}
\end{equation}
A trivial observation is that the N{\'e}el state 
on the necklace  with $N$ sites is generated by the first
order recurrence  relation,

\begin{eqnarray}
&|{\rm Neel},2 N+1 > = |\downarrow> \;\;|{\rm Neel}, 2N > & \label{a4} \\
&|{\rm Neel}, 2 N> = |T_{\uparrow}> \; |{\rm Neel}, 2N-1> & \nonumber 
\end{eqnarray}
Quantum 
fluctuations around  the N{\'e}el state amount to local changes of the 
form (see Fig. 27)

\begin{eqnarray}
& |\downarrow> \; |T_{\uparrow}> \longrightarrow 
 |(\downarrow, T_{\uparrow})> \equiv 
 |\uparrow> \; |T_{0}> & \nonumber \\
&  |T_{\uparrow}> \; |\downarrow> \longrightarrow 
 |(T_{\uparrow}, \downarrow)> \equiv 
 |T_{0}> \; |\uparrow> & \label{a5} \\
& |\downarrow> \; |T_{\uparrow}> \; |\downarrow> 
\longrightarrow 
 |(\downarrow, T_{\uparrow},\downarrow)> \equiv 
 |\uparrow> \; |T_{\downarrow}> \;|\uparrow> & \nonumber 
\end{eqnarray}

\noindent
Hence a globally perturbed N{\'e}el state is a 
coherent superposition  of states of the form,

\begin{equation}
\downarrow \; (T_{\uparrow} 
\downarrow)  (T_{\uparrow} \; \downarrow) \; T_{\uparrow} \;
 (\downarrow \;  T_{\uparrow}
\downarrow) \; (T_{\uparrow} 
\downarrow)  T_{\uparrow} \; \downarrow \; (T_{\uparrow} \;
 \downarrow) \;  T_{\uparrow}
\downarrow 
\label{a6}
\end{equation}

\noindent
where the parenthesis denote the quantum fluctuations
given in (\ref{a5}). The RVA state is a linear superposition
of states of the type (\ref{a6}) weighted with probability 
amplitudes which are the variational parameters.

The RVA state is generated by the following recursion relations (RR),

\begin{equation}
\begin{array}{rl}
|2 N+1 > = &  |\downarrow> \;\;|2N > \\
&+ u \;|(\downarrow, \;T_{\uparrow})> 
\;|2N-1 > \\
&+v\; |(\downarrow, T_{\uparrow}, \downarrow)> |2N-2 >  \\
|2 N> = & |T_{\uparrow}> \; |2N-1> \\
& + u \; |(T_{\uparrow},\downarrow)> \; ;|2N-2 > 
\end{array} 
\label{a7}
\end{equation}

\noindent
with the initial condition

\begin{equation}
|N=1> \equiv |\downarrow>, \;\;\;|N=0>\equiv 1
\label{a8}
\end{equation}

\noindent
To compute the energy of the state $|N>$ we define the
following matrix elements,

\begin{eqnarray}
& Z_N = <N|N> & \label{a9} \\
& E_N = <N| H_N |N> & \nonumber 
\end{eqnarray}

\noindent
where $H_N$ is the Hamiltonian of the system with $N$ sites.

The RR's for the states (\ref{a7}) imply
a set of recursion relations for the matrix elements (\ref{a9}).
Form the norm we get,

\begin{eqnarray}
& Z_{2N+1} = Z_{2N} + u^2 Z_{2N-1} + v^2 Z_{2N-2} & \label{a10} \\
& Z_{2N} = Z_{2N-1} + u^2 Z_{2N-2} & \nonumber  
\end{eqnarray}

\noindent The initial conditions are,

\begin{equation}
Z_{0} =  Z_{1} = 1
\label{a11}
\end{equation}

\noindent
The RR's for the energy are,

\begin{eqnarray}
\begin{array}{rl}
E_{2N+1} = & E_{2N} + u^2 E_{2N-1} + v^2 E_{2N-2} 
 + \\&
J(\sqrt{2} u - {1\over 2}) Z_{2N-1} + 
J(2\sqrt{2} uv - v^2) Z_{2N-2} \\&
+ {J\over 2} v^2 Z_{2N-3}\label{a12} \\
E_{2 N} = & E_{2N-1} + u^2 E_{2N-2} + 
\\&
 J(\sqrt{2} u - {1\over 2}) Z_{2N-2} +
J u^2 Z_{2N-3} {J\over 2} v^2 Z_{2N-4}
\end{array}
\nonumber 
\end{eqnarray}

\noindent
while in this case the initial data are

\begin{equation}
E_0 = E_1 = 0 \label{a13}
\end{equation}

Minimizing the GS energy in the limit $N >>1$ we find that
the GS per plaquette is given by  -0.4822 J, which corresponds
to an energy per site of the associated spin chain equal
to -0.7233 J. The values of the variational parameters are
given by $u=-0.3288 $ and $v=0.1691$.

\subsection{Case $x= 1/3$}

The most probable state for this doping is 
given by (see Fig. 7)

\begin{equation}
|\circ> \; |S> \; |\circ> \; |S> \; \cdots
|\circ> \; |S> \; |\circ> 
\label{a14}
\end{equation}

\noindent
Analogously to Eq.(\ref{a5}) we define 
local fluctuations around (\ref{a14})
in terms of the states 
$|(\circ,S)>, |(S,\circ)>$ and $|(\circ,S,\circ)>$ depicted
in Fig. 27.
The $x=1/3$ RVA state can then be constructed from the following
RR's ( see Fig. 17)

\begin{equation}
\begin{array}{rl}
|2 N+1 > =  & |\circ> \;\;| 2N > \\
&+ a \;|(\circ, \;S)> 
\;|2N-1 > \\&
+ b\; |(\circ, S, \circ)> |2N-2 > \\
|2 N> = & |S> \; |2N-1> \\
& + a \; |(S,\circ)> \;| 2N-2 > 
\end{array}
\label{a15} 
\end{equation}

The norm of $|N>$ satisfies the RR's (\ref{a10}) with
the replacements $u \rightarrow a, v \rightarrow b$. 
The RR's for the energy matrix elements are given by,

\begin{eqnarray}
& E_{2N+1} = E_{2N} + a^2 E_{2N-1} + b^2 E_{2N-2} & \label{a16} \\
& + (\sqrt{2} (-2t)a - J a^2) Z_{2N-1} + \sqrt{2} (-2t)2ab Z_{2N-2} &
\nonumber \\
& + a^2 (-J/4) (a^2 Z_{2N-3} + b^2 Z_{2N-4}) & \nonumber \\
& + b^2 (-J/4) (2 Z_{2N-3} + a^2 Z_{2N-4}) & \nonumber \\
& E_{2N} = E_{2N-1} + a^2 E_{2N-2} 
- (2 \sqrt{2} t a +J a^2) Z_{2N-2} & \nonumber  \\
& (-J/4) (a^2 Z_{2N-3} + 2b^2 Z_{2N-4}) &  \nonumber \\ 
& + a^2 (-J/4) (2 Z_{2N-3} + a^2 Z_{2N-4}) & \nonumber 
\end{eqnarray}

\noindent
The initial conditions for both $Z_N$ and $E_N$ are 
the same as for the undoped case.
In the limit $N \rightarrow \infty$ we find

\begin{equation}
\lim_{N \rightarrow \infty} \frac{E_0(N)}{N} = -0.7387,
\;\;\;a = 0.7873 ,\;\;\; b = 0.5478 \label{aa16}
\end{equation}

\noindent 
which give the asymptotic values of the curves in Fig. 18.

\begin{figure}
\epsfxsize=9cm \epsffile{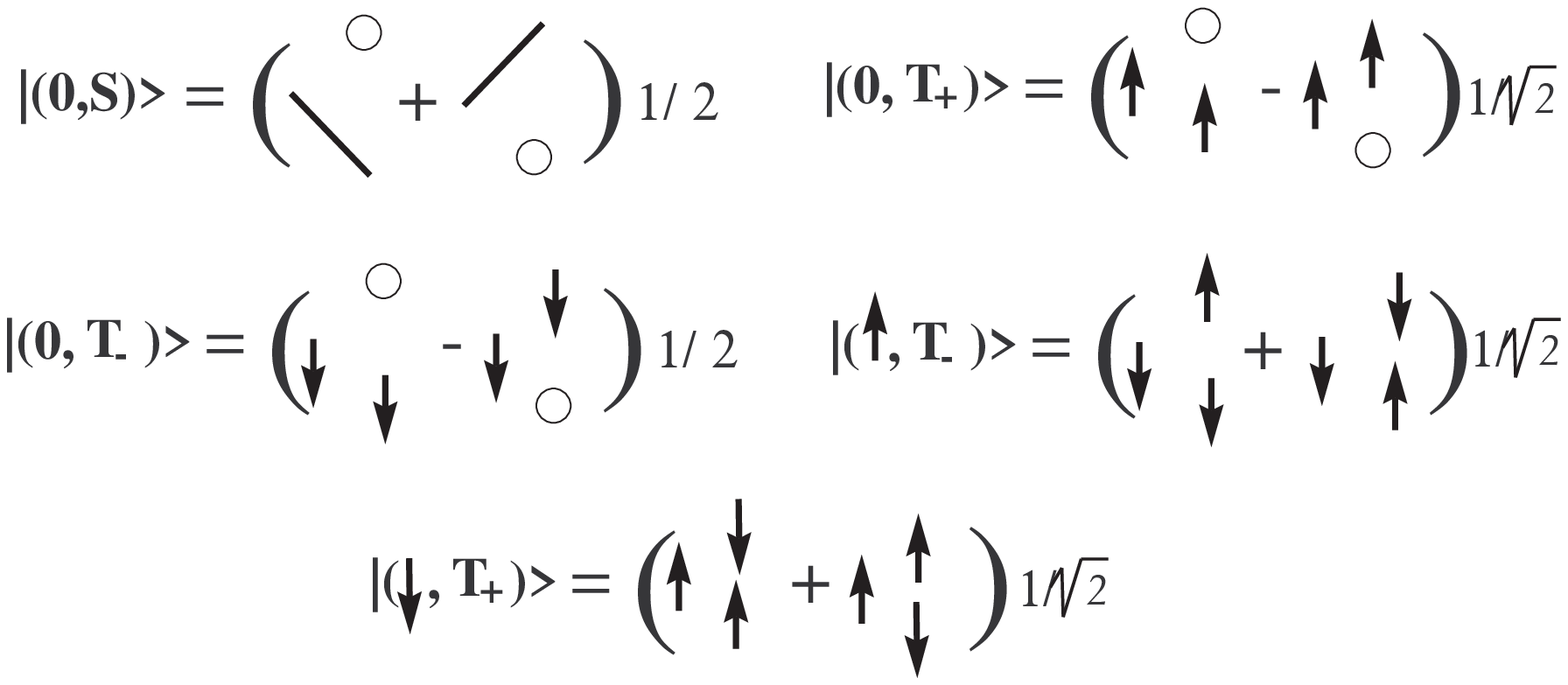}
\narrowtext
\caption[]{ A pictorical representation of fluctuation states as constructed
in text for the RVA method in the necklace ladder.
Here $T_+$ and $T_-$ stand for $T_{\uparrow}$ and $T_{\downarrow}$,
respectively.}
\label{fig27a} 
\end{figure}
\noindent

\subsection{ Cases $0 < x < 1/3$}

In the underdoped region $0 < x < 1/3$ 
we have observed with the DMRG that many of the GS
that one gets, and particularly those listed in Table
2 can be understood as quantum fluctuations 
around a ``classical" state $|\psi_0>$. 
This state has the generic structure already seen in the cases
$x=0$ and 1/3 (see (\ref{a6}) and (\ref{a14})), namely

\begin{equation}
|\psi_0\rangle = |l_{2N+1}\rangle |l_{2N}\rangle \cdots 
|l_{2}\rangle |l_{1}\rangle 
\label{a17}
\end{equation}

\noindent where the states $|l_{i}\rangle$, $i=1,2, \cdots 2N,2N+1$,
are taken to be

\begin{equation}
|l_{odd}\rangle = \{ |0\rangle, |\downarrow \rangle, |\uparrow \rangle
\} = \{ 1, 3, 5 \}
\label{a18}
\end{equation}

\begin{equation}
|l_{even}\rangle = \{ |S\rangle, |T_{\uparrow} \rangle, 
|T_{\downarrow} \rangle \} = \{2, 4, 6 \}
\label{a19}
\end{equation}

\noindent
Notice that we do not allow the holes on  the minor diagonals
of the classical state $|\psi_0>$. They will go there after considering
the fluctuations \cite{comment}. 
Based on the DMRG results as well as physical 
considerations, we shall
allow the following  pairs $|l_i > | l_{i+1}>$ in $|\psi_0>$

\begin{equation}
|0\rangle|S\rangle , \; |0\rangle|T_{\uparrow}\rangle , \;
|\downarrow \rangle|T_{\uparrow}\rangle , \;
|0\rangle|T_{\downarrow}\rangle , \; |\uparrow \rangle|T_{\downarrow}\rangle 
\label{a20}
\end{equation}

\begin{equation}
|S\rangle |0\rangle, \; |T_{\uparrow}\rangle |0\rangle, \;
|T_{\uparrow}\rangle |\downarrow \rangle, \;
|T_{\downarrow}\rangle |0\rangle, \;|T_{\downarrow}\rangle |\uparrow \rangle
\label{a21}
\end{equation}

\begin{figure}
\epsfxsize=9cm \epsffile{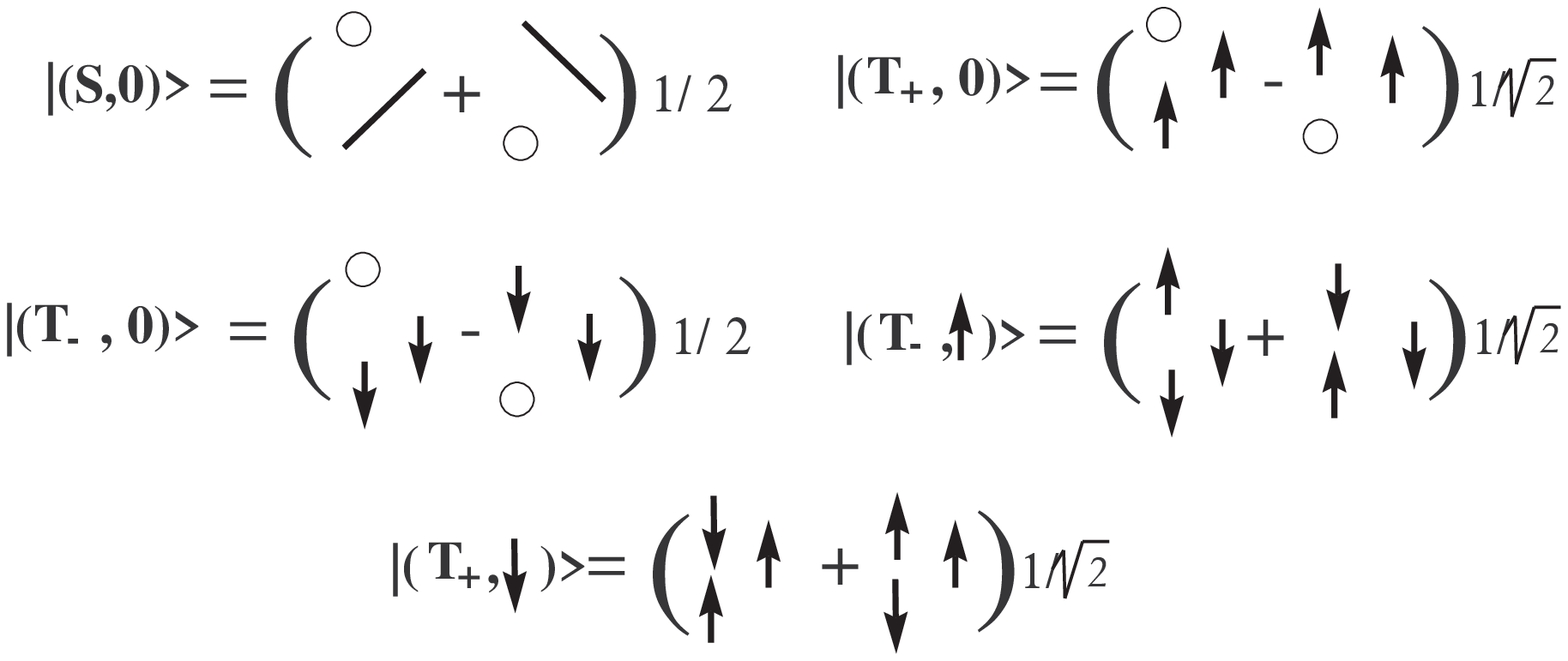}
\narrowtext
\caption[]{ A pictorical representation of fluctuation states as constructed
in text for the RVA method in the necklace ladder.
Here $T_+$ and $T_-$ stand for $T_{\uparrow}$ and $T_{\downarrow}$,
respectively.}
\label{fig27b} 
\end{figure}
\noindent

\begin{figure}
\epsfxsize=9cm \epsffile{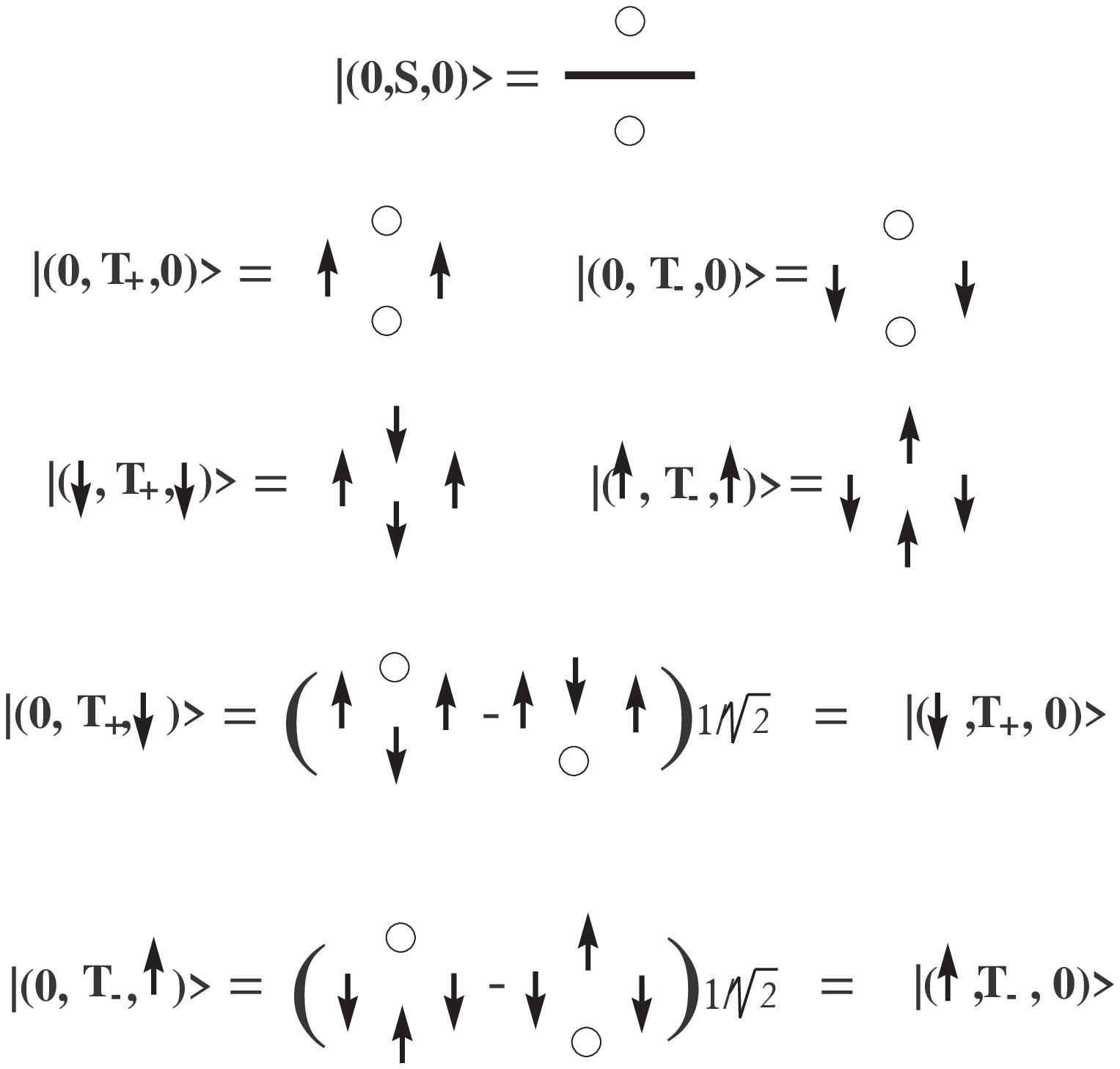}
\narrowtext
\caption[]{ A pictorical representation of fluctuation states as constructed
in text for the RVA method in the necklace ladder.
Here $T_+$ and $T_-$ stand for $T_{\uparrow}$ and $T_{\downarrow}$,
respectively.}
\label{fig27c} 
\end{figure}
\noindent

\noindent
This connectivity of the states making 
up a certain $|\psi_0\rangle$
state can be summarized 
in a graph  in which we place a site for each and
every 6 states in  (\ref{a18}), (\ref{a19}), 
and joint them  by links
whenever it is possible to 
find them one next to each other in the $|\psi_0\rangle$
state according to the allowed local configurations (\ref{a20},\ref{a21}).
This graph is depicted  in Fig. 30,  
and coincides with  
the Dynkin diagram of the exceptional Lie Group $E_6$.

Now we can characterize 
every admissible classical state $|\psi_0\rangle$
in a geometrical fashion: 
each $|\psi_0\rangle$ is a path in the so called 
Bratelli diagram associated to the Dynkin diagram of $E_6$.
This Bratelli diagram is shown in
Fig. 31. 
The way it is constructed 
is apparent in that figure: one starts with the 3 possible
site-states (\ref{a18}) 
located one on top of each other. 
These states are located by the label
$l_1$ of the first site of the diagonal ladder.
Then we link them to the 3 possible 
rung-states (\ref{a19}) 
according to the connectivity prescribed in Fig. 30.
These rung-states are located by the 
label $l_2$ of the second position of the diagonal ladder.
Once this is achieved, 
the rest of the graph in Fig. 31 
is made up by reflecting this 
basic piece over the rest of the labels $l_3, l_4, \ldots,l_{2N},
l_{2N+1}$. 
Observe that the $x=0$ and $x=1/3$ states
discussed previously correspond to straight paths of the 
Bratelli diagram (\ref{29}). 
A similar 
type of construction is also used in Statistical
Mechanics in the context of  the Face Models\cite{Bax}. 

\begin{figure}
\epsfxsize=9cm \epsffile{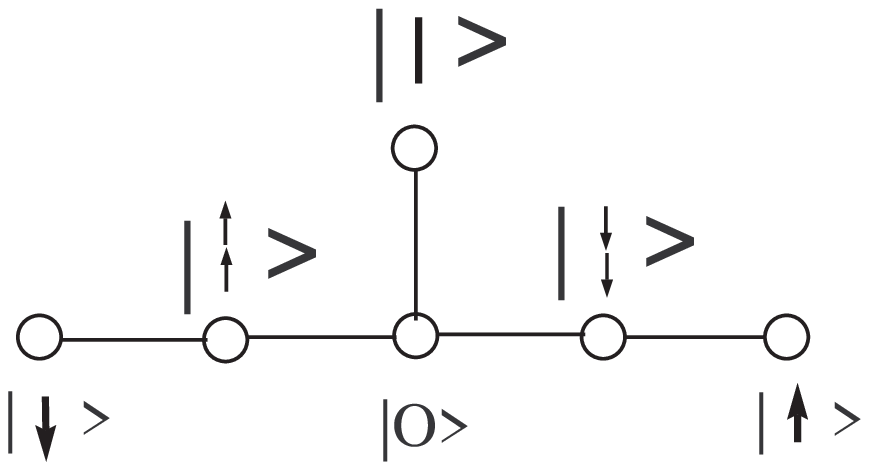}
\narrowtext
\caption[]{Dynkin diagram of the exceptional Lie group $E_6$ and its associated
site and rung states contributing to the variational RVA method for the 
necklace ladder. These 6 states make up every classical state $|\psi_0\rangle$
on the underdoped ladder according to the connectivity of this diagram.}
\label{fig28} 
\end{figure}
\noindent

The quantum fluctuations around $|\psi_0>$ amounts
to considering the normalized 
states $|(l_i,l_{i+1})>$ and
$|(l_i,l_{i+1},l_{i+2})>$ depicted  in Figs. 27-29.
An interesting property of these states is that they are
orthogonal, i.e.

\begin{equation}
\langle l_j|(l_i,l_{i+1},l_{i+2})\rangle = 0, \;\; j=i, i+1, i+2
\label{a22}
\end{equation}

\begin{equation}
\langle (l_j,l_{j+1})|(l_i,l_{i+1},l_{i+2})\rangle = 0,\; \; j=i, i+1
\label{a23}
\end{equation}

\noindent
The RVA state built upon $|\psi_0>$ is generated by the RR's,
(see Fig. 30)

\begin{eqnarray}
\begin{array}{rl}
|2N+1\rangle = &
|l_{2N+1}\rangle |2N\rangle + a_N |(l_{2N+1},l_{2N})\rangle |2N-1\rangle
\\
&+ b_N |(l_{2N+1},l_{2N},l_{2N-1})\rangle |2N-2\rangle \\&\\

|2N\rangle = & |l_{2N}\rangle  |2N-1\rangle 
+ c_N |(l_{2N},l_{2N-1})\rangle |2N-2\rangle 
\end{array}
\label{a25}
\end{eqnarray}

\noindent provided with the initial data,

\begin{equation}
|1\rangle \equiv |l_1\rangle,\;  |0\rangle \equiv 1
\label{a26}
\end{equation}

\noindent
Using the orthogonality conditions (\ref{a22},\ref{a23})
it is easy to get the  RR's satisfied by the norm of the RVA state

\begin{equation}
Z_{2N+1} = Z_{2N} + a_N^2 Z_{2N-1} + b_N^2 Z_{2N-2}
\label{a27}
\end{equation}

\begin{equation}
Z_{2N} = Z_{2N-1} + c_N^2 Z_{2N-2}
\label{a28}
\end{equation}

\begin{figure}
\epsfysize=6cm 
\epsfxsize=9cm \epsffile{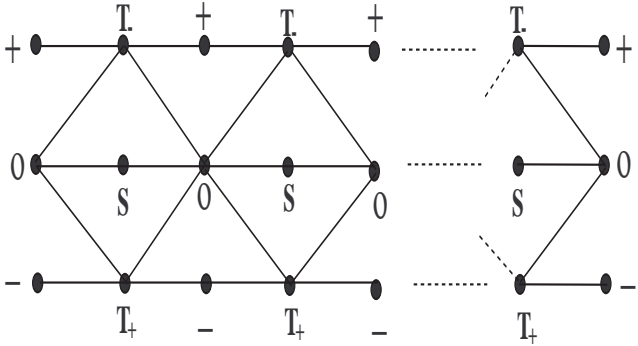}
\narrowtext
\caption[]{Bratelli diagram of the exceptional Lie group $E_6$.
It serves to classify all the classical states $\psi_0$ appearing in
the RVA treatment of the underdoped necklace ladder: every path on this
diagram characterizes one of those classical states and provides its 
quantum numbers. Here  $T_+$ and $T_-$ stand for $T_{\uparrow}$ and $T_{\downarrow}$,
respectively. Also, $S$ denotes a singlet state and $O$ represents a hole.}
\label{fig28} 
\end{figure}
\noindent

\noindent
The RR's for the energy have the structure given below

\[
E_{2N+1} = E_{2N} + a_N^2 E_{2N-1} + b_N^2 E_{2N-2} +
\]
\[
Z_{2N-1} (\epsilon_{2N+1,2N}^{(0)} + a_N^2 \epsilon_{2N+1,2N}^{(2)}
\]
\[ 
+ 2a_N \epsilon_{2N+1,2N}^{(1)}) +
\]
\[
Z_{2N-2} (c_N^2 \epsilon_{2N+1,2N,2N-1}^{(4)} 
+ 
\]
\[
a_N^2 \bar{\epsilon}_{2N+1,2N,2N-1}^{(4)} +
b_N^2 \epsilon_{2N+1,2N,2N-1}^{(8)} + 
\]
\[
2b_N c_N \epsilon_{2N+1,2N,2N-1}^{(3)} 
+ 2a_N b_N \bar{\epsilon}_{2N+1,2N,2N-1}^{(3)}) +
\]
\[
Z_{2N-3} (a_N^2 a_{N-1}^2 \epsilon_{2N+1,2N,2N-1,2N-2}^{(5)} + 
\]
\[
b_N^2 \bar{\epsilon}_{2N+1,2N,2N-1,2N-2}^{(6)}) +
\]
\[
Z_{2N-4} (a_N^2 b_{N-1}^2 \epsilon_{2N+1,2N,2N-1,2N-2,2N-3}^{(7)} + 
\]
\begin{equation}
b_N^2 c_{N-1}^2 \bar{\epsilon}_{2N+1,2N,2N-1,2N-2,2N-3}^{(7)}) 
\label{aa28}
\end{equation}

\[
E_{2N} = E_{2N-1} + c_N^2 E_{2N-2} 
\]
\[+ Z_{2N-2} (\epsilon_{2N,2N-1}^{(0)} + c_N^2 \epsilon_{2N,2N-1}^{(2)} +
2c_N \epsilon_{2N,2N-1}^{(1)}) +
\]
\[
Z_{2N-3} (a_{N-1}^2 
\epsilon_{2N,2N-1,2N-2}^{(4)} 
\]
\[
+ c_{N}^2 \bar{\epsilon}_{2N,2N-1,2N-2}^{(4)}) +
\]
\[
Z_{2N-4} (b_{N-1}^2 \epsilon_{2N,2N-1,2N-2,2N-3}^{(6)} 
\]
\begin{equation}
+ c_{N}^2 c_{N-1}^2 \epsilon_{2N,2N-1,2N-2,2N-3}^{(5)})
\label{a299}
\end{equation}

\noindent The $\epsilon $ symbols 
in Eqs. (\ref{a299}) are reduced energy matrix
elements involving the states defined in Fig. 27-29. 
To simplify the RR's for the energy
we adopt the notation: $\epsilon_{2N+1,2N,\dots} \equiv
\epsilon_{l_{2N+1}, l_{2N}, \dots} $, where $l_{2N+1},
l_{2N}, \dots $ run over the six possible states, $1,\dots,6$,  of 
the Dynkin diagram $E_6$. All the non vanishing
values of $\epsilon$ are shown in tables 3.

\begin{figure}
\epsfysize=6cm 
\epsfxsize=9cm \epsffile{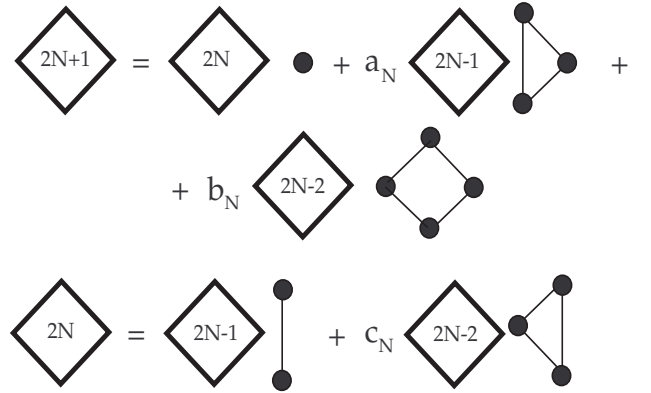}
\narrowtext
\caption[]{A pictorical representation of the Recursion Relations
in Eq.
(\ref{a25}) employed to generate the variational states in the 
RVA treatment of the underdoped ($0 \leq x \leq 1/3$) necklace ladder.
Here $a$, $b$ and $c$ are local variational parameters.
A square denotes a bulk state on a ladder of a length given by its number inside.
The black circles and solid lines represent generic fluctuation states as 
explained in the text.}
\label{fig30} 
\end{figure}
\noindent

Given the previous equations we set up 
the following strategy 
to derive  the results presented in Section V.

\begin{itemize}

\item Fix the length $N$ of the ladder,
the number  of holes $h$ and the third component of the spin $S^z$
of the whole ladder.

\item  Generate all the $|\psi_0>$ 

configurations with those quantum 
numbers $N,h,S^z$. Generically the number
of configurations grows exponentially. For
example the 7 plaquette case studied in Section V has 
N=15 and  a total of $3 \times10^4$
configurations.

\item  Compute the energy of the state associated with the 
zero-order state $|\psi_0>$ 
using the recursion relations and find the 
variational parameters which lead to a minimum energy. For example for
the 7 plaquette ladder we used 21 independent variational
parameters.

\item Extract the state $|\psi_0>$ which has the absolute minimum
energy for a given $N,h$ and $S^z$.

\end{itemize}

\begin{center}
\begin{tabular}{|c|c|c|}
\hline
$\epsilon_{i_1,i_2}^{(0)}$ & (3,4), (5,6),(4,3),(6,5)  & $-J$  \\
\hline 
$\epsilon_{i_1,i_2}^{(1)}$ & (1,2),(1,4),(1,6),(2,1),(4,1),(6,1)
 & $-\sqrt{2} t$\\
\hline 
$\epsilon_{i_1,i_2}^{(1)}$ & (3,4),(5,6),(4,3),(6,5)
 & ${J\over \sqrt{2}}$\\
\hline 
$\epsilon_{i_1,i_2}^{(2)}$ & (1,2),(2,1) & $-J$ \\
\hline 
$\epsilon_{i_1,i_2}^{(2)}$ & (3,4),(5,6),(4,3),(6,5)& $-{J\over 2}$ \\
\hline
\end{tabular}
\end{center}
\begin{center}
Table 3 a) Non-vanishing reduced energy matrix elements for
$\epsilon_{i_1,i_2}^{(0)}=
\langle l_{i_2}| \langle l_{i_1}|h^{(tJ)}_{i_1 i_2}|l_{i_1} \rangle|l_{i_2} \rangle$  
where the tJ Hamiltonian acts only on the states specified by its subscripts.
Likewise for the elements $\epsilon_{i_1,i_2}^{(1)}=
\langle (l_{i_1}l_{i_2})|h^{(tJ)}_{i_1 i_2}|l_{i_1} \rangle|l_{i_2} \rangle$,
$\epsilon_{i_1,i_2}^{(2)}=
\langle (l_{i_1}l_{i_2})|h^{(tJ)}_{i_1 i_2}|(l_{i_1},l_{i_2})\rangle$.
\end{center}


\begin{center}
\begin{tabular}{|c|c|c|}
\hline
$\epsilon_{i_1,i_2,i_3}^{(3)}$ & (1,2,1) & $-\sqrt{2} t$   \\
\hline 
$\epsilon_{i_1,i_2,i_3}^{(3)}$ & (1,4,3),(1,6,5) & $-t$ \\
\hline
$\epsilon_{i_1,i_2,i_3}^{(3)}$ & (3,4,1),(5,6,1) & ${-J\over 2}$ \\
\hline
$\epsilon_{i_1,i_2,i_3}^{(3)}$ & (3,4,3),(5,6,5) & ${J\over \sqrt{2}}$ \\
\hline
$\bar{\epsilon}_{i_1,i_2,i_3}^{(3)}$ & (1,2,1) & $-\sqrt{2} t$ \\
\hline
$\bar{\epsilon}_{i_1,i_2,i_3}^{(3)}$ & (1,4,3),(1,6,5) & ${J\over 2}$ \\
\hline
$\bar{\epsilon}_{i_1,i_2,i_3}^{(3)}$ & (3,4,1),(5,6,1) & $t$ \\
\hline
$\bar{\epsilon}_{i_1,i_2,i_3}^{(3)}$ & (3,4,3),(5,6,5) & ${J\over \sqrt{2}}$ \\
\hline
\end{tabular}
\end{center}
\begin{center}
Table 3 b) Non-vanishing reduced energy matrix elements for
$\epsilon_{i_1,i_2,i_3}^{(3)}=
\langle (l_{i_1},l_{i_2},l_{i_3})|h^{(tJ)}_{i_1 i_2}
|l_{i_1} \rangle|(l_{i_2},l_{i_3}) \rangle$  
where the tJ Hamiltonian acts only on the states specified by its subscripts.
Likewise for the elements $\bar{\epsilon}_{i_1,i_2,i_3}^{(3)}=
\langle (l_{i_1},l_{i_2},l_{i_3})|h^{(tJ)}_{i_2 i_3}|
(l_{i_1},l_{i_2}) \rangle|l_{i_3} \rangle$.
\end{center}


\begin{center}
\begin{tabular}{|c|c|c|}
\hline
$\epsilon_{i_1,i_2,i_3}^{(4)}$ & (2,1,2),(2,1,4),(2,1,6),(4,1,2),(6,1,2)&$-{J\over 2}$\\
\hline
$\epsilon_{i_1,i_2,i_3}^{(4)}$&(3,4,3),(5,6,5),(3,4,1),(5,6,1) & -${J\over 2}$ \\
\hline
$\epsilon_{i_1,i_2,i_3}^{(4)}$ & (6,1,4),(4,1,6) & $J$   \\
\hline
$\bar{\epsilon}_{i_1,i_2,i_3}^{(4)}$ & (2,1,2),(2,1,4),(2,1,6),(4,1,2),(6,1,2) & $-{J\over 2}$ 
\\
\hline
$\bar{\epsilon}_{i_1,i_2,i_3}^{(4)}$ & (3,4,3),(5,6,5),(1,4,3),(1,6,5) & $-{J\over 2}$  \\
\hline
$\bar{\epsilon}_{i_1,i_2,i_3}^{(4)}$ & (4,1,6),(6,1,4) & $J$ \\ 
\hline
\end{tabular}
\end{center}
\begin{center}
Table 3 c) Non-vanishing reduced energy matrix elements for
$\epsilon_{i_1,i_2,i_3}^{(4)}=
\langle (l_{i_2},l_{i_3})| \langle l_{i_1}|h^{(tJ)}_{i_1 i_2}
|l_{i_1} \rangle|(l_{i_2},l_{i_3}) \rangle$  
where the tJ Hamiltonian acts only on the states specified by its subscripts.
Likewise for the elements $\bar{\epsilon}_{i_1,i_2,i_3}^{(4)}=
\langle l_{i_3}| \langle (l_{i_1},l_{i_2})|h^{(tJ)}_{i_2 i_3}|
(l_{i_1},l_{i_2}) \rangle|l_{i_3} \rangle$. 
\end{center}


\begin{center}
\begin{tabular}{|c|c|c|}
\hline
$\epsilon_{i_1,i_2,i_3}^{(8)}$ & (3,4,3),(5,6,5) & $-2J$ \\
\hline
$\epsilon_{i_1,i_2,i_3}^{(8)}$ & (1,4,3),(1,6,5),(3,4,1),(5,6,1) & $-J$\\
\hline
\end{tabular}
\end{center}
\begin{center}
Table 3 d) Non-vanishing energy matrix elements
$\epsilon_{i_1,i_2,i_3}^{(8)}=
\langle (l_{i_1},l_{i_2},l_{i_3})|(h^{(tJ)}_{i_1 i_2}+h^{(tJ)}_{i_2 i_3})
|(l_{i_1},l_{i_2},l_{i_3}) \rangle$  
where the tJ Hamiltonian acts only on the states specified by its subscripts.
. 
\end{center}


\begin{center}
\begin{tabular}{|c|c|c|}
\hline
$\epsilon_{i_1,i_2,i_3,i_4}^{(5)}$ &     (1,2,1,2),(1,2,1,4),(1,2,1,6),(1,4,1,2),(1,6,1,2) & 
$-{J\over 4}$ \\
\hline
$\epsilon_{i_1,i_2,i_3,i_4}^{(5)}$ &     (1,4,1,6),(1,6,1,4),(3,4,3,4),(5,6,5,6),(3,4,1,2) & 
$-{J\over 2}$ \\
\hline
$\epsilon_{i_1,i_2,i_3,i_4}^{(5)}$ & 
(5,6,1,2),(3,4,1,6),(5,6,1,4),(3,4,1,4),(5,6,1,6) & $-{J\over 2}$ \\
\hline
$\epsilon_{i_1,i_2,i_3,i_4}^{(5)}$ &     (2,1,2,1),(2,1,4,1),(2,1,6,1),(4,1,2,1),(6,1,2,1) & 
$-{J\over 4}$ \\
\hline
$\epsilon_{i_1,i_2,i_3,i_4}^{(5)}$ &     (2,1,4,3),(2,1,6,5),(4,3,4,3),(6,5,6,5),(4,1,4,3) & 
$-{J\over 2}$ \\
\hline
$\epsilon_{i_1,i_2,i_3,i_4}^{(5)}$ & 
(6,1,6,5),(4,1,6,1),(6,1,4,1),(4,1,6,5),(6,1,4,3) & $-{J\over 2}$ \\
\hline
\end{tabular}
\end{center}
\begin{center}
Table 3 e) Non-vanishing energy matrix elements 
$\epsilon_{i_1,i_2,i_3,i_4}^{(5)}=
\langle (l_{i_3},l_{i_4})| \langle (l_{i_1},l_{i_2})|h^{(tJ)}_{i_2 i_3}
|(l_{i_1},l_{i_2}) \rangle|(l_{i_3},l_{i_4}) \rangle$  
where the tJ Hamiltonian acts only on the states specified by its subscripts. 
\end{center}


\begin{center}
\begin{tabular}{|c|c|c|}
\hline
$\epsilon_{i_1,i_2,i_3,i_4}^{(6)}$ &    (2,1,2,1),(2,1,4,1),(2,1,6,1),(2,1,4,3),(2,1,6,5) & 
$-{J\over 2}$ \\
\hline
$\epsilon_{i_1,i_2,i_3,i_4}^{(6)}$ & 
(4,1,2,1),(6,1,2,1),(4,1,6,5),(6,1,4,3) & $-{J\over 2}$ \\
\hline
$\epsilon_{i_1,i_2,i_3,i_4}^{(6)}$ & (4,1,6,1),(6,1,4,1) & $-J$ \\
\hline
\end{tabular}
\end{center}
\begin{center}
Table 3 f) Non-vanishing energy matrix elements 
$\epsilon_{i_1,i_2,i_3,i_4}^{(6)}=
\langle (l_{i_2},l_{i_3},l_{i_4})| \langle l_{i_1}|h^{(tJ)}_{i_1 i_2}
|l_{i_1} \rangle|(l_{i_2},l_{i_3},l_{i_4}) \rangle$  
where the tJ Hamiltonian acts only on the states specified by its subscripts. 
\end{center}


\begin{center}
\begin{tabular}{|c|c|c|}
\hline
$\bar{\epsilon}_{i_1,i_2,i_3,i_4}^{(6)}$ &  (1,2,1,2),(1,2,1,4),(1,2,1,6),(1,4,1,2),(1,6,1,2) & 
$-{J\over 2}$ \\
\hline
$\bar{\epsilon}_{i_1,i_2,i_3,i_4}^{(6)}$ & 
(3,4,1,2),(5,6,1,2),(3,4,1,6),(5,6,1,4) & $-{J\over 2}$ \\
\hline
$\bar{\epsilon}_{i_1,i_2,i_3,i_4}^{(6)}$ & (1,4,1,6),(1,6,1,4) & $-J$  \\
\hline
\end{tabular}
\end{center}
\begin{center}
Table 3 g) Non-vanishing energy matrix elements 
$\bar{\epsilon}_{i_1,i_2,i_3,i_4}^{(6)}=
\langle l_{i_4}|\langle (l_{i_1},l_{i_2},l_{i_3})|h^{(tJ)}_{i_3 i_4}
|(l_{i_1},l_{i_2},l_{i_3}) \rangle |l_{i_4} \rangle$  
where the tJ Hamiltonian acts only on the states specified by its subscripts. 
\end{center}


\begin{center}
\begin{tabular}{|c|c|c|}
\hline
$\epsilon_{i_1,i_2,i_3,i_4,i_5}^{(7)}$ &   (1,2,1,2,1),(1,2,1,4,1),(1,2,1,6,1),(1,2,1,4,3) & 
$-{J\over 4}$ \\
\hline
$\epsilon_{i_1,i_2,i_3,i_4,i_5}^{(7)}$ & 
(1,2,1,6,5),(1,4,1,2,1),(1,6,1,2,1) & $-{J\over 4}$ \\
\hline
$\epsilon_{i_1,i_2,i_3,i_4,i_5}^{(7)}$ & 
(1,4,1,6,1),(1,6,1,4,1),(1,4,1,6,5),(1,6,1,4,3) & $-{J\over 2}$ \\
\hline
$\epsilon_{i_1,i_2,i_3,i_4,i_5}^{(7)}$ & 
(3,4,3,4,3),(5,6,5,6,5),(3,4,3,4,1),(5,6,5,6,1) & $-{J\over 2}$ \\
\hline
$\epsilon_{i_1,i_2,i_3,i_4,i_5}^{(7)}$ & 
(3,4,1,2,1),(5,6,1,2,1),(3,4,1,4,3),(5,6,1,6,5) & $-{J\over 2}$ \\
\hline
$\epsilon_{i_1,i_2,i_3,i_4,i_5}^{(7)}$ & 
(3,4,1,4,1),(5,6,1,6,1),(3,4,1,6,5) & $-{J\over 2}$ \\
\hline
$\epsilon_{i_1,i_2,i_3,i_4,i_5}^{(7)}$ & 
(5,6,1,4,3),(3,4,1,6,1),(5,6,1,4,1) & $-{J\over 2}$ \\
\hline
\end{tabular}
\end{center}
\begin{center}
Table 3 h) Non-vanishing reduced energy matrix elements for
$\epsilon_{i_1,i_2,i_3,i_4,i_5}^{(7)}=
\langle (l_{i_3},l_{i_4},l_{i_5})| \langle (l_{i_1},l_{i_2})|h^{(tJ)}_{i_2 i_3}
|(l_{i_1},l_{i_2}) \rangle|(l_{i_3},l_{i_4},l_{i_5}) \rangle$  
where the tJ Hamiltonian acts only on the states specified by its subscripts.
\end{center}


\begin{center}
\begin{tabular}{|c|c|c|}
\hline
$\bar{\epsilon}_{i_1,i_2,i_3,i_4,i_5}^{(7)}$  &  (1,2,1,2,1),(1,4,1,2,1),(1,6,1,2,1),(3,4,1,2,1) 
& $-{J\over 4}$ \\
\hline
$\bar{\epsilon}_{i_1,i_2,i_3,i_4,i_5}^{(7)}$  & 
(5,6,1,2,1),(1,2,1,4,1),(1,2,1,6,1) & $-{J\over 4}$ \\
\hline
$\bar{\epsilon}_{i_1,i_2,i_3,i_4,i_5}^{(7)}$  &  (1,6,1,4,1),(1,4,1,6,1),(5,6,1,4,1),(3,4,1,6,1) 
& $-{J\over 2}$ \\
\hline
$\bar{\epsilon}_{i_1,i_2,i_3,i_4,i_5}^{(7)}$& 
(3,4,3,4,3),(5,6,5,6,5),(1,4,3,4,3),(1,6,5,6,5) & $-{J\over 2}$ \\
\hline
$\bar{\epsilon}_{i_1,i_2,i_3,i_4,i_5}^{(7)}$  & (1,2,1,4,3),(1,2,1,6,5),(3,4,1,4,3),(5,6,1,6,5) 
& $-{J\over 2}$ \\
\hline
$\bar{\epsilon}_{i_1,i_2,i_3,i_4,i_5}^{(7)}$  & 
(1,4,1,4,3),(1,6,1,6,5),(5,6,1,4,3) & $-{J\over 2}$ \\
\hline
$\bar{\epsilon}_{i_1,i_2,i_3,i_4,i_5}^{(7)}$  & 
(3,4,1,6,5),(1,6,1,4,3),(1,4,1,6,5) & $-{J\over 2}$ \\
\hline
\end{tabular}
\end{center}
\begin{center}
Table 3 i) Non-vanishing reduced energy matrix elements for
$\bar{\epsilon}_{i_1,i_2,i_3,i_4,i_5}^{(7)}=
\langle (l_{i_4},l_{i_5})| \langle (l_{i_1},l_{i_2},l_{i_3})| h^{(tJ)}_{i_3 i_4}
|(l_{i_1},l_{i_2},l_{i_3}) \rangle |(l_{i_4},l_{i_5}) \rangle$  
where the tJ Hamiltonian acts only on the states specified by its subscripts. 
\end{center}

\subsection{Electronic Density G.S. Expectation Values}

Here we derive the general equations to compute the electronic density
expectation values using the RVA method. In particular, we have plotted
in Fig. 8 the results for the doping case $x=1/3$ in the necklace ladder
and compared it with the DMRG results showing a good agreement. 

Let us denote the density electronic operator at the site position $R$
as $d_R$, namely,

\begin{equation}
d_R = n_{R,\uparrow} + n_{R,\downarrow}
\label{x1}
\end{equation}

\noindent where $n_{R,\uparrow}$ and $n_{R,\downarrow}$ are the number 
operators for fermions with spins up and down, respectively.

\noindent For a diagonal ladder of length $N$ (sites+rungs) we need to
compute the V.E.V. of the density operator $d_R$ in the ground state,
which we denote as,

\begin{equation}
d_N(R) = \langle N |d_R| N\rangle
\label{x2}
\end{equation}

\noindent When the position $R$ is even we shall need an additional index
$k=1,2$ to locate the upper site ($k=1$) and the lower site ($k=2$) of the
rung.

\noindent Notice that these density V.E.V. (\ref{x2}) are not normalized. 
We may introduce normalized densities as,

\begin{equation}
\hat{d}_N(R) = {\langle N |d_R| N\rangle \over \langle N|N\rangle } =
{d_N(R)\over Z_N}
\label{x2b}
\end{equation}

The densities $\hat{d}_N(R)$ takes on values from 0 to 1 depending 
on whether we
find a hole with maximum probability ($\hat{d}_N(R)=0$) 
or one electron ($\hat{d}_N(R)=1$).

Using the RR's for the diagonal (\ref{a25}) ladder we may find also RR's for the 
unnormalized  V.E.V.'s: 

\begin{equation}
d_{2N+1}(R) = d_{2N}(R) + a_N^2 d_{2N-1}(R) + b_N^2 d_{2N-2}(R)
\label{x3}
\end{equation}

\begin{equation}
d_{2N}(R) = d_{2N-1}(R) + c_N^2 d_{2N-2}(R) 
\label{x4}
\end{equation}

\noindent whenerver the position $R$ of the insertion is not near the end
of the diagonal ladder state. The derivation of these
RR's follow closely that of the norms $Z's$ using the orthogonality relations.

Now we need to determine the boundary or initial conditions to feed those
RR's. Let us consider the cases $R$ odd and even separately.

\subsubsection{$R$ Odd}

As for the odd sites the RR's of the states are third order,
we have to compute directly the values $d_R(R)$, $d_{R+1}(R)$ and $d_{R+2}(R)$.
Clearly we also have,

\begin{equation}
d_{R-1}(R) = 0
\label{x5}
\end{equation}

\noindent Now, as $R$ is odd, let us set $R=2r+1$. Then, using the RR's
we find,

\begin{equation}
d_{2r+1}(2r+1) = \delta_{2r+1}^{(o)} Z_{2r} + a_r^2 Z_{2r-1} + b_r^2 Z_{2r-2}
\label{x6}
\end{equation}

\noindent In the derivation of these RR's we need to compute the following
reduced density matrix elements,

\begin{equation}
\delta_{2r+1}^{(o)} \equiv \langle l_{2r+1}|d_{2r+1}|l_{2r+1}\rangle = 
\left\{ \begin{array}{ccc}
0 & \mbox{if} & l_{2r+1}=|\circ\rangle \\
1 & \mbox{if} & l_{2r+1}=\{ |\uparrow\rangle, |\downarrow\rangle\}
\end{array} \right.
\label{x7}
\end{equation}

\begin{equation}
\delta_{2r+1,2r}^{(o)} \equiv 
\langle (l_{2r+1},l_{2r})|d_{2r+1}|(l_{2r+1},l_{2r})\rangle = 1
\label{x8}
\end{equation}

\[
\delta_{2r+1,2r,2r-1}^{(o)} \equiv 
\]
\begin{equation}
\langle (l_{2r+1},l_{2r},l_{2r-1})|d_{2r+1}|(l_{2r+1},l_{2r},l_{2r-1})\rangle = 1
\label{x9}
\end{equation}

Likewise we proceed with the initial value $d_{R+1}(R)$ and we find,

\begin{equation}
d_{2r+2}(2r+1) = d_{2r+1}(2r+1) + c_{r+1}^2 Z_{2r} 
\label{x10}
\end{equation}

\noindent where we have used the result,

\begin{equation}
\langle (l_{2r+2},l_{2r+1})|d_{2r+1}|(l_{2r+2},l_{2r+1})\rangle = 1
\label{x11}
\end{equation}

Finally, we proceed similarly with the other initial value $d_{R+2}(R)$ and we find,

\[
d_{2r+3}(2r+1) = d_{2r+2}(2r+1) + 
\]
\begin{equation}
a_{r+1}^2 d_{2r+1}(2r+1) + b_{r+1}^2 Z_{2r}
\label{x11b}
\end{equation}

\subsubsection{$R$ Even}

As for the even positions (rungs) the RR's (\ref{a25}) of the states are second order,
we have to compute directly the values $d_{R}(R,k)$ and $d_{R+1}(R,k)$
($k=1,2$). Clearly
we also have,

\begin{equation}
d_{R-1}(R,k) = 0
\label{x5b}
\end{equation}

\noindent Now, as $R$ is even, let us set $R=2r$. Then, using the RR's
we find,

\begin{equation}
d_{2r}(2r,k) =  Z_{2r-1} + c_r^2 \delta_{2r,2r-1,k}^{(e)} Z_{2r-2} 
\label{x12}
\end{equation}

\noindent In the derivation of these RR's we need to compute the following
reduced density matrix elements,

\begin{equation}
\delta_{2r,k}^{(e)} \equiv \langle l_{2r}|d_{2r,k}|l_{2r}\rangle = 1
\label{x13}
\end{equation}

\begin{equation}
\delta_{2r,2r-1,k}^{(e)} \equiv 
\langle (l_{2r},l_{2r-1})|d_{2r,k}|(l_{2r},l_{2r-1})\rangle 
\label{x14}
\end{equation}

Using the expressions of the fluctuation states $|(l_{2r},l_{2r-1})\rangle$ we find 
that the only nonvanishing elements are as follows,

\begin{equation}
\begin{array}{ccc}
\delta_{1,2;k}^{(e)} = {1\over 2}, &
\delta_{1,4;k}^{(e)} = {1\over 2}, &
\delta_{1,6;k}^{(e)} = {1\over 2} 
\end{array}
\label{x15}
\end{equation}

\begin{equation}
\begin{array}{ccc}
\delta_{2,1;k}^{(e)} = {1\over 2},  &
\delta_{4,1;k}^{(e)} = {1\over 2},  &
\delta_{6,1;k}^{(e)} = {1\over 2} 
\end{array}
\label{x16}
\end{equation}

\begin{equation}
\begin{array}{cc}
\delta_{3,4;k}^{(e)} = 1,  &
\delta_{5,6;k}^{(e)} = 1  
\end{array}
\label{x17}
\end{equation}

\begin{equation}
\begin{array}{cc}
\delta_{4,3;k}^{(e)} = 1,  &
\delta_{6,5;k}^{(e)} = 1  
\end{array}
\label{x18}
\end{equation}

\noindent where we are following the notation for the site/rung states,
$\{ 1,3,5 \} = \{ |\circ\rangle, |\downarrow \rangle, |\uparrow \rangle \}$,
$\{ 2,4,6 \} = \{ |S \rangle, |T_{\uparrow} \rangle, |T_{\downarrow} \rangle  \}$.

Likewise we proceed with the other initial value $d_{R+1}(R)$ and we find,

\[
d_{2r+1}(2r,k) =  d_{2r}(2r,k) + 
\]
\begin{equation}
a_r^2 \delta_{2r+1,2r,k}^{(e)} Z_{2r-1} +
b_r^2 \delta_{2r+1,2r,2r-1,k}^{(e)} Z_{2r-2} 
\label{x19}
\end{equation}

\noindent In the derivation of these RR's we need to compute the following
reduced density matrix elements,

\begin{equation}
\delta_{2r+1,2r,2r-1,k}^{(e)} \equiv 
\langle (l_{2r+1},l_{2r},l_{2r-1})|d_{2r,k}|(l_{2r+1},l_{2r},l_{2r-1})\rangle 
\label{x20}
\end{equation}

Using the expressions of the fluctuation states $|(l_{2r+1},l_{2r},l_{2r-1})\rangle$ we find 
that the only nonvanishing elements are as follows,

\begin{equation}
\begin{array}{cc}
\delta_{3,4,3;k}^{(e)} = 1,  &
\delta_{5,6,5;k}^{(e)} = 1  
\end{array}
\label{x21}
\end{equation}

\begin{equation}
\begin{array}{cc}
\delta_{3,4,1;k}^{(e)} = {1\over 2},  &
\delta_{5,6,1;k}^{(e)} = {1\over 2}  
\end{array}
\label{x22}
\end{equation}

\begin{equation}
\begin{array}{cc}
\delta_{1,4,3;k}^{(e)} = {1\over 2},  &
\delta_{1,5,6;k}^{(e)} = {1\over 2}  
\end{array}
\label{x23}
\end{equation}

\section*{Appendix B: Spectrum of the $\lowercase{t-J}$ model on the plaquette}

The $t-J$ Hamiltonian on the plaquette has the symmetry group $D_4$
of the square. This implies that the eigenstates can be classified
with the irreps of $D_4$, together with the number of holes $h$
and the total spin $S$. $D_4$ has 5 irreps, four of which
are one-dimensional  $A_1,B_1, A_2, B_2$  and one two-dimensional
irrep $E$. From the character table of this group one sees that
the irreps $A_1, B_1$ have, in our terminology,
 even parity in both diagonals, while the irreps $A_2, B_2$ are
odd. The parity of the states in the irrep $E$ is $(+,-)$
and $(-,+)$ for the two diagonals.
In Table 3 we show the analytic expression of the energy
and the value for the cases $J=0$ and $J=0.5$ ($t=1)$.
In Fig. 33 we show a plot with all the energies as functions
of $J/t$. Observe the crossover between the lowest energy states
at $J/t \sim 1.37$.

\begin{center}
\begin{tabular}{|c|c|c|c|c|c|}
\hline
$h$ & $S$ & $D_4$ & Energy & $J=0$ & $J=0.5$ \\
\hline 
0 & 0 & $B_2$ & $- 3 J$ & 0 &$ - 1.5$ \\
0 & 1 & $A_2$ & $- 2 J$ & 0 &$ -1$ \\
0 & 0 & $A_1$ & $- J$ & 0 &$ -0.5$ \\
0 & 1 & $E$ &$ - J$ & 0 & $-0.5$ \\
0 & 2 & $B_2$ &   0 & 0 &  0   \\
\hline
1 & 1/2 & $E$ & $- J - \sqrt{ 3 t^2 +J^2/4}$ &$ - \sqrt{3}$ &$ - 2.25$ \\
1 & 1/2 & $B_2$ & $- 3J/2 - t$ & $-1$ & $- 1.75$ \\
1 & 1/2 & $B_1$ & $-J/2 - t$ &$ -1$ &$ -1.25$ \\
1 & 1/2 & $A_2$ & $-3J/2 +t$ & 1 & 0.25 \\
1 & 1/2 & $A_1$ & $-J/2 + t$ & 1 & 0.75 \\
1 & 1/2 & $E$ &$ -J + \sqrt{3 t^2 + J^2/4}$ & $\sqrt{3}$ & 1.25 \\
1 & 3/2 & $A_2$ & $- 2 t$ & $-2$ & $-2$ \\
1 & 3/2 & $E$ & 0 & 0 & 0 \\
1 & 3/2 & $B_2$ & $2t$ & 2 & 2 \\
\hline
2 & 0 & $A_1$ & $-J/2 -\sqrt{8 t^2 + J^2/4}$ &$ -\sqrt{8}$ & $-3.09$ \\
2 & 0 & $B_1$ &$0$ & 0 & $0$ \\
2 & 0 & $E,B_2$ &$ -J$ & 0 & $-0.5$ \\
2 & 0 & $A_1$ & $-J/2 + \sqrt{8 t^2 + J^2/4}$ & $\sqrt{8} $ & 2.59 \\
2 & 1 & $E$ & $- 2 t$ & $-2$ & $-2$ \\
2 & 1 & $A_2,B_1$ & $0$ & 0 & 0 \\
2 & 1 & $E$ & $2 t$ & 2 & 2 \\
\hline
3 & 1/2 & $A_1$ & $- 2 t$ & $-2$ &$ -2$ \\
3 & 1/2 & $E$ & $0$ & 0 & 0 \\
3 & 1/2 & $B_1$ & $2 t$ & 2 & 2 \\
\hline
4 & 0 & $A_1$ & 0 & 0 & 0 \\
\hline
\end{tabular}
\end{center}
\begin{center}
Table 4. Here $h$ is the number of holes, $S$ the total spin,
The third column denotes the irrep of the 
$D_4$ group. We give the values of the energy for $J=0$ and
0.5 ( $t=1$).
\end{center}

\begin{figure}
\epsfysize=10cm 
\epsfxsize=9cm \epsffile{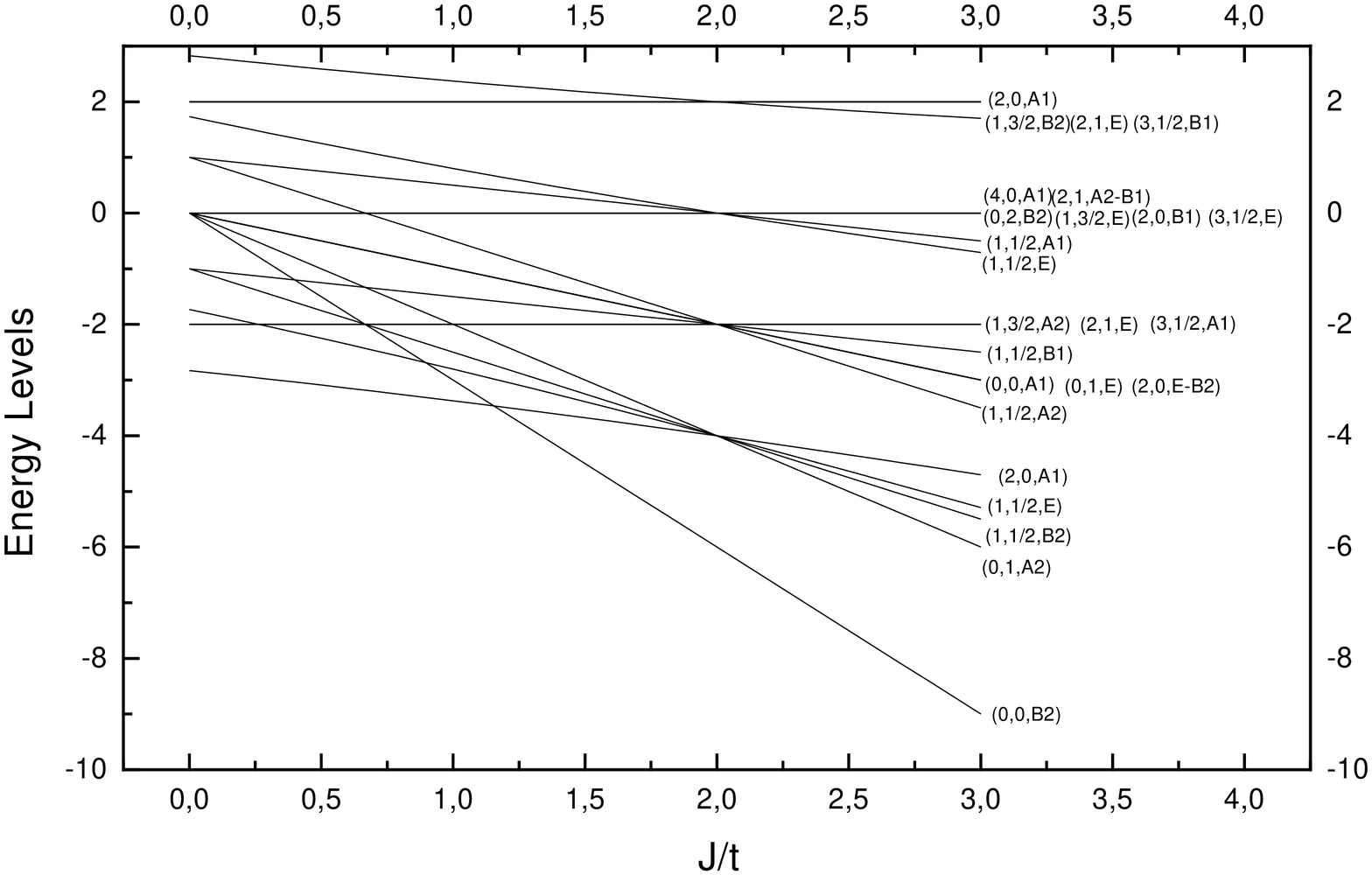}
\narrowtext
\caption[]{Plot showing the energy 
levels for a $tJ$ model on the $2\times 2$ plaquette
as a function of the coupling $J/t$.
The energy levels are classified according the $D_4$
symmetry group of the square as shown in table 4.}
\label{fig31} 
\end{figure}
\noindent

\section*{Appendix C: Plaquette derivation of the equivalence
between the Haldane state and the RVB state of the 2-leg
spin ladder}

There has been some debate in the past as to whether
the Haldane state of the spin 1 chain is in the same
phase as the GS of the  AFH 2-leg ladder. 
The general consensus is that they both belong to the same
universality class characterized by a spin gap, finite
spin correlation length and non vanishing string order parameter.
The DMRG study of Ref. \cite{mapping} demonstrated a continuous
mapping between these systems, and pointed out the equivalence
between a dimer-RVB state on a composite spin model (which is
a ladder model with some extra hopping terms) and the AKLT model.

\begin{figure}
\epsfxsize=9cm \epsffile{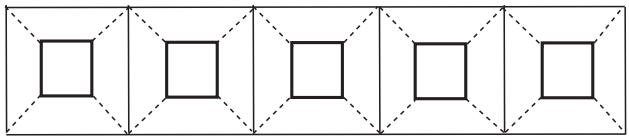}
\narrowtext
\caption[]{Plaquette construction (small interior squares plus
dashed lines) of the rectangular ladder
with $n_l=2$ legs.}
\label{fig32} 
\end{figure}
\noindent

This suggests that there must be a direct way to relate
the valence bond construction of the spin 1 AKLT state\cite{AKLT} 
and the dimer-RVB picture of the 2-leg ladder\cite{WNS,RVA1}. 
We shall show that this is indeed possible through
the plaquette construction of the 2-leg ladder, which
is shown diagrammatically in Figs. 34 and 35. 

In Fig. 34 the 2-leg ladder is split into plaquettes 
connected by two links. We have generalized somewhat
the medial construction in this case, since two plaquettes
are allowed to have more than one common link. 
The interesting point is that the filling factor
$x$ of the ladder and the one of the extended model, $x_p$,
are related in exactly the same manner as the 2D lattices 
(see Eq.(\ref{34})). Thus  $x=0$ corresponds to $x_p=1/2$.
Let us assume for the moment that each plaquette has spin 1.
Then the effective interaction describing the coupling of these
spins will be an AFH model. This suggests that we can construct a valence
bond state to approximate the plaquette ground state by
drawing nearest neighbors bonds among the elementary spins
between plaquettes, as shown in the upper part of Fig. 35. Now,
if we project out of this state any $S=0$ components of each plaquette,
we get the AKLT state, where each plaquette is a pure spin $S=1$.
If, instead, we Gutzwiller project the $t_d$ links, as shown in
Fig. 35, we get
the dimer-RVB state proposed in references\cite{WNS,RVA1}. 

\begin{figure}
\epsfxsize=9cm \epsffile{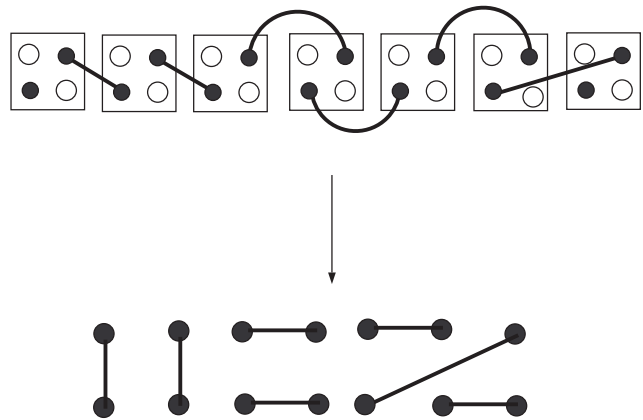}
\narrowtext
\caption[]{(Below) A dimer-RVB state on a rectangular 2-leg ladder
obtained as the projection (above) of a doped $x_p=1/2$ valence bond state on the
decorated (dual) lattice associated to the rectangular ladder.}
\label{fig33} 
\end{figure}
\noindent

Hence, the plaquette model acts as intermediate
system, for which different projections generate either the AKLT
state of the spin-1 chain or the dimer-RVB state of the two-leg
ladder.


\begin{thebibliography}{99}
\bibitem{hubbard} The link Hamiltonian (\ref{14}) can be written
in terms of the bonding and antibonding operators $c_{B/A,s}
= (c_{L,s} \pm c_{R,s})/\sqrt{2}$ as 
$h_{\langle L,R \rangle} = 2 t_d \; c^\dagger_{A,s} \; c_{A,s}$. 
Hence the 4 states  $ |0\rangle, c_{B,s}^\dagger |0 \rangle $
and $c_{B,\uparrow}^\dagger \; 
c_{B,\downarrow}^\dagger |0 \rangle$  have zero energy and
are in one-to-one correspondence with those of a Hubbard model
while the remaining 12 states, which contain antibonding operators,  
decouple in the limit $t_d \rightarrow \infty$.
\bibitem{comment} This election of states to construct the ground state in
the RVA method depends on the coupling regime under consideration.
Here we are assuming $J/t = 0.35$ throughout this paper so that kinetics 
effects are stronger that exchange effects. Were we in a regime with 
$J>t$, then the selection of building states would be different.
The physics of the problem dictates the way in which the RVA method is set up.
\bibitem{comment2} The $Cu O_2$ planes of the cuprates yield another example of medial
construction. Let ${\cal L}_{\rm Cu}$ and ${\cal L}_{\rm O}$ 
denote the square lattices formed
by the $Cu$ and $O$ ions respectively. Then the $O$ lattice
is the medial graph of the $Cu$ lattice, i.e. ${\cal L}_{\rm O}
= {\cal M}( {\cal L}_{\rm Cu})$. A plaquette corresponds in this
case to a single $Cu$ ion surrounded by 4 Oxigens. This implies that
the 3 band Hubbard model admits a natural $t_d$-extension.
\end{thebibliography}
\end{document}